\documentclass[9pt,preprint2]{aastex}
\usepackage{multirow}
\usepackage{multicol}

\usepackage{natbib}
\usepackage{epsfig}
\usepackage{pdflscape}

\newcommand\arcpt{${{\lower3pt\hbox{$^{\prime\prime}$}}\atop{\raise4pt\hbox{.}}}$}
\newcommand\msun{$M_\odot$}

\slugcomment{submitted to the {\it Astronomical Journal} 27 August 2013}

\shorttitle{45 Nearby Low-Mass Young Systems} \shortauthors{Riedel et al.}

\begin{document}

\title{The Solar Neighborhood. XXXIII. Parallax Results from the CTIOPI
0.9m Program: Trigonometric Parallaxes of Nearby Low-Mass Active and
Young Systems}

\author{Adric~R.~Riedel$^{1,2}$\altaffilmark{a}, Charlie~T.~Finch$^{3}$\altaffilmark{a}, Todd~J.~Henry$^{4}$\altaffilmark{a}, John~P.~Subasavage$^{5}$\altaffilmark{a}, Wei-Chun~Jao$^{4}$\altaffilmark{a}, Lison~Malo$^{6}$, David~R.~Rodriguez$^{7}$ Russel~J.~White$^{4}$, Douglas~R.~Gies$^{4}$,  Sergio~B.~Dieterich$^{4}$\altaffilmark{a}, Jennifer~G.~Winters$^{4}$\altaffilmark{a}, Cassy~L.~Davison$^{4}$\altaffilmark{a}, Edmund~P.~Nelan$^{8}$, Sarah~C.~Blunt$^{9,2}$, Kelle~L.~Cruz$^{1,2}$,Emily~L.~Rice$^{9,2}$,Philip A.~Ianna$^{10}$\altaffilmark{a}}

\affil {$^{1}$Department of Physics and Astronomy, Hunter College, The City University of New York, 695 Park Avenue, New York, NY 10065}
\affil {$^{2}$Department of Astrophysics, American Museum of Natural History, Central Park West at 79th Street, New York, NY 10024}
\affil {$^{3}$Astrometry Department, U.S. Naval Observatory, Washington DC 20392}
\affil {$^{4}$United States Naval Observatory, Flagstaff, AZ 86001, USA}
\affil {$^{5}$Department of Physics and Astronomy, Georgia State University, P.O. Box 5060, Atlanta, GA 30302-5060}
\affil {$^{6}$D{\' e}partement de Physique et Observatoire du Mont-Megantic, Universit{\' e} de Montr{\' e}al, C.P. 6128, Succursale Centre-Ville, Montr{\' e}al, QC, Canada H3C 3J7}
\affil {$^{7}$Departamento de Astronomia, Universidad de Chile, Casilla 36-D, Las Condes, Santiago, Chile}
\affil {$^{8}$Space Telescope Science Institute}
\affil {$^{9}$Department of Engineering Science and Physics, College of Staten Island, 2800 Victory Boulevard, New York, NY 10314}
\affil {$^{10}$Department of Astronomy, University of Virginia, Charlottesville, VA 22904}

\email{ar494@hunter.cuny.edu}


\altaffiltext{a}{Visiting Astronomer, Cerro Tololo Inter-American
Observatory.  CTIO is operated by AURA, Inc.\ under contract to the
National Science Foundation.}

\vfil\eject

\begin{abstract}

We present basic observational data and association membership analysis for 45
young and active low-mass stellar systems from the ongoing RECONS photometry 
and astrometry program at the Cerro Tololo Inter-American Observatory.  Most 
of these systems have saturated X-ray emission ($\log{\frac{L_{X}}{L_{bol}}} > -3.5$) 
based on X-ray fluxes from the ROSAT All-Sky Survey, and many are significantly 
more luminous than main-sequence stars of comparable color.  
We present parallaxes and proper motions, Johnson-Kron-Cousins $VRI$ 
photometry, and multiplicity observations from the CTIOPI program on the 
CTIO 0.9m telescope. To this we add low-resolution optical spectroscopy 
and line measurements from the CTIO 1.5m telescope, 
and interferometric binary measurements from the Hubble Space
Telescope Fine Guidance Sensors.  We also incorporate data from published 
sources: $JHK_S$ photometry from the 2MASS
point source catalog; X-ray data from the ROSAT All-Sky Survey; and
radial velocities from literature sources. Within the sample of 45 systems, 
we identify 21 candidate low-mass pre-main-sequence members of nearby
associations, including members of $\beta$ Pictoris, TW Hydrae,
Argus, AB Doradus, two ambiguous $\approx$30 Myr old
systems, and one object that may be a member of the Ursa Major moving group.  
Of the 21 candidate young systems, 14 are newly identified as a result of 
this work, and six of those are within 25 parsecs of the Sun.


\end{abstract}

\section{Introduction}
\label{sec:intro}
Over the past 20 years, a variety of loose associations 
have been discovered, with distances ($<$100 pc) much closer than any 
star-forming region, and ages ($\sim$100 Myr) much younger than
any comparably close moving group (e.g. Ursa Major, \citealt{King2003}).  
These associations include such well-studied groups as TW Hydra
\citep{de-la-Reza1989,Gregorio-Hetem1992}, $\beta$ Pictoris
\citep{Barrado-y-Navascues1999}, Tucana-Horologium
\citep{Zuckerman2001}, Argus \citep{Torres2003}, AB Doradus
\citep{Zuckerman2004a}, Carina \citep{Torres2008}, and 
Columba \citep{Torres2008}.

Most of the currently known members of these associations are
solar-type or hotter stars, reflecting a bias toward bright stars that 
are surveyed in the proper motion catalogs, {\it HIPPARCOS} and {\it TYCHO-2}. 
We are likely left without information on most of the members of these 
associations; for instance, M dwarfs make up {\it at least} 75\% of 
all nearby stars \citep{Henry2006}, but make up less than half of the known 
members of young associations (see Table \ref{tab:statisticstable}).

The dearth of M dwarfs is a distinct issue with star formation theory, 
and presents difficulties with our understanding of young associations. 
M dwarfs, because of their lower masses, should be more easily scattered by 
dynamical interactions than solar-type stars, and thus the current spatial 
and kinematic boundaries 
of the associations will not necessarily contain many of the associated 
stars.  By virtue of numbers, they will better inform the Initial Mass 
Function (IMF) measurements of young associations, which currently appear to 
be very different from the field IMF \citep{Schlieder2011}.  M dwarfs 
provide an advantage for exoplanet research because they are redder and 
dimmer, which enhances the contrast between the stars and any forming 
planets in their star systems.  Finally, M dwarfs take significantly longer
to reach the main sequence (at least 200 Myr, \citealt{Dotter2008}), 
which makes it easier to identify and obtain precise ages for young M dwarfs.

To address the issue of missing M dwarfs, we present the results of a new 
survey of young and active M dwarfs, as part of the Research Consortium On 
Nearby Stars\footnote{http://www.recons.org} (RECONS) exploration of 
the Solar Neighborhood.  We present 45 nearby star systems 
with M dwarf primaries (35 with new astrometry and photometry) observed 
during the Cerro Tololo Inter-American Obseratory Parallax Investigation 
(CTIOPI). Our analysis of youth is based on absolute trigonometric parallaxes, 
Johnson-Kron-Cousins $VRI$ photometry, spectral types, variability, 
kinematic analyses, and surface gravity estimates.  The systems discussed 
herein include both known and new candidate pre-main-sequence members of 
the $\beta$ Pictoris, TW Hydra, Tucana-Horologium, Columba, Argus, AB Doradus, 
and Castor associations.

Identifying young stars (specifically, pre-main-sequence young stars) is a 
complicated process.  There are many signatures of youth 
that can be detected in M dwarf stars.  Unfortunately, 
there is no single indicator that completely describes youth, and none of the 
parameters are foolproof.  Lithium equivalent widths (EWs), spectral accretion 
signatures, and protoplanetary disks are all only found in young stars, but they 
are short-lived effects, and stars can lack those properties and still be 
pre-main-sequence objects.  Conversely, the other parameters (overluminosity, 
low surface gravity, chromospheric activity) are long-lived in M dwarfs and 
the presence of that signature does not necessarily 
mean the star is young -- particularly, most forms of stellar activity can also 
be induced by magnetic interactions with a close binary.  Our analysis must 
therefore use multiple independent lines of evidence to identify young stars, 
similar to recent efforts by \citep{Shkolnik2009,Shkolnik2012}.

The key to our present analysis is trigonometric parallaxes: With parallaxes, 
we have significantly improved constraints on the kinematics of the systems, 
AND (along with our photometry) we can use the positions of the constituent 
stars on an HR diagram with confidence to determine if a system falls along a 
particular association's isochrones. With our low resolution spectroscopic 
data, we can measure spectroscopic features sensitive to surface gravity.  
We can also measure activity features, though the latter are less useful 
indicators for M dwarfs.

In section \ref{sec:sample}, we discuss the sample selection.  In
section \ref{sec:data}, we discuss the observations and
reductions of the data in this paper.  Section \ref{sec:analysis}
describes the methods we used to determine the ages and association 
memberships of these objects, and in section \ref{sec:results} we 
discuss the outcome of the youth analysis carried out on our stars.

This is the thirteenth paper publishing parallax results from the ongoing CTIOPI 
program\footnote{A complete table of all published parallaxes is available here: 
http://www.recons.org/} at the CTIO 0.9m telescope.

\section{The Sample}
\label{sec:sample}
    The sample of 45 star systems in this paper was drawn from the hundreds of 
    targets in the CTIOPI parallax target list. CTIOPI, by virtue of its 
    location, is limited to objects at declinations south of $+$30 degrees.  
    By using the Tektronix imager at the CTIO 0.9m telescope, CTIOPI is further 
    limited to stars between 9$^{th}$ and 18$^{th}$ magnitudes in at least one 
    of our three Johnson-Kron-Cousins $VRI$ filters.  CTIOPI generally targets 
    M dwarf stars whose estimated 
    distances -- either from literature, or from our own photometric 
    distance estimates \citep{Hambly2004,Henry2004} -- place them 
    within 25 pc of the Sun.  The targets in this paper are consequently 
    all nearby bright M dwarfs (see Table \ref{tab:photometry}, Column 16).

    From the CTIOPI target list, we identified potentially young stars 
    using X-ray saturation ($\log{\frac{L_{X}}{L_{bol}}} > -3.5$) 
    as an indicator of chromospheric activity, and overluminosity -- herein 
    defined as being more than one magnitude brighter than a single main-sequence 
    star of comparable colors -- as an indicator of low surface gravity.  As shown by 
    \citet{Zuckerman2004}, X-ray emission in M dwarfs is saturated and remains 
    at the $\log{\frac{L_{X}}{L_{bol}}} \approx -3.0$ level in stars at
    least as old as the Hyades (600 Myr), the oldest of the associations we consider 
    here. Therefore, the presence of saturation-level X-ray emission is an 
    excellent indicator (though not guarantor) of youth.

    Accordingly, X-ray photometry was obtained from the ROSAT All-Sky Survey
    (RASS; \citealt{Voges1999,Voges2000}) for every available star on the
    CTIOPI parallax program; the resulting systems have
    X-ray detections with better than 25\%~errors on the counts and are within 
    25\arcsec~(95\% detection radius, \citealt{Voges1999}) 
    of the proper-motion-corrected epoch 1991, equinox J2000 coordinates 
    (close to the mean epoch of RASS). The calculation of
    $\log{\frac{L_{X}}{L_{bol}}}$ is taken from
    \citet{Schmitt1995}, using bolometric calculations from
    \citet{Casagrande2008}.  We found positive evidence of saturated X-ray 
    emission for 39 of our star systems, many of which are also overluminous.

\begin{figure}
\center
\includegraphics[angle=90,width=.6\textwidth]{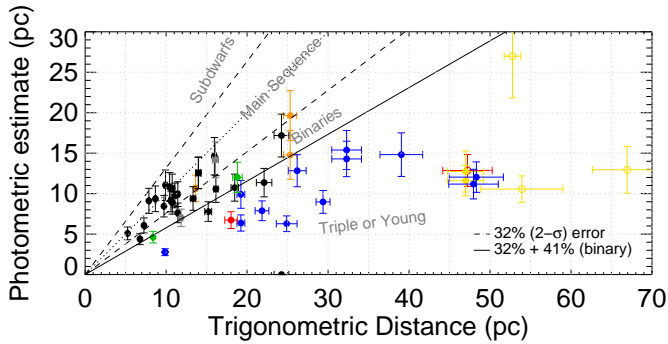}
\caption{A diagram of the 51 resolved components of the 45 systems in this 
paper, demonstrating their overluminosity in terms of their trigonometric 
distances (X axis) and photometric distances (Y axis).  Objects are 
color-coded by the association to which they are ultimately linked as members:
TW Hydrae are shown in yellow; $\beta$ Pictoris in blue; Tucana-Horologium, 
Columba, and Carina in red; Argus in green; AB Doradus in orange; 
and Castor and Ursa Major in gray.  Open circles have no RASS detection.
\label{fig:photometric}}
\end{figure}

    The remaining six star systems came to our attention purely by the overluminosiy criterion.
    They exhibit no X-ray emission in the ROSAT catalogs, but their luminosity
    makes it difficult to explain them as unresolved binaries or triples.
    (Figure \ref{fig:photometric}). 

    Among the 45 systems considered here, we have individual photometry
    and astrometry of 51 components\footnote{As seen in Tables 
    \ref{tab:photometry} and \ref{tab:astrometry}, GJ 2022AC was observed 
    for standard photometry but not astrometry, and GJ 799B has resolved
    astrometry but not photometry}, because six of our star systems contain 
    binaries with separations more than 1 arcsecond.
    Many of the stars in this paper were
    originally identified as active by \citet{Riaz2006}, and several have
    already been identified as young by \citet{Zuckerman2004}, 
    \citet{Shkolnik2012} and \citet{Malo2013}.   
    Ten systems were published 
    in previous papers in this series; their astrometry and photometry 
    is reprinted from the earlier papers without change.

\section{Observations and Reductions}
\label{sec:data}
    \subsection{Photometry}
    \label{sec:photometry}
        All CTIOPI photometry is conducted with the CTIO 0.9m telescope,
        initially (1999-2003) under the NOAO Survey Programs grant;
        later (2003-present) via the SMARTS
        Consortium.  Photometry is conducted in three filters (Tektronix 
        \#2 $VRI$), utilizing only the central quarter 
        (6.8$\times$6.8\arcmin~FOV, 401 mas pixel$^{-1}$) of the 
        Tektronics 2048x2046 CCD to minimize distortions for astrometry.  
        These values are then transformed to standard $V_J R_{KC} 
        I_{KC}$\footnote{Subscripts: ``J'' indicates Johnson, and 
          ``KC'' indicates Kron-Cousins (SAAO system), 
          which is more often known as Cousins.  The central wavelengths
          for $V_J, R_{KC}$, and $I_{KC}$ are 5475, 6425, and 8075 \AA,
          respectively.} (hereafter without subscripts) photometry
        using observations of standards from \citet{Graham1982},
        \citet{Landolt1992}, and \citet{Landolt2007}.  The resulting
        photometry can be found in Table \ref{tab:photometry}.
        Further details of the observation and reduction procedures can be
        found in \citet{Jao2005} and \citet{Winters2011}.  The photometric
        errors quoted in Columns 3, 4, and 5 of Table \ref{tab:photometry}
        combine the Poisson errors, errors on the nightly calibration
        fit, and standard deviation of multiple nights of photometry
        (see Column 6).  Generally, the latter is the greatest
        contributor to the collective error, particularly when the star 
        is active, as these stars are.  This $VRI$ photometry, along with 
        2MASS $JHK$ photometry \citep{Cutri2003}, is printed in Columns 
        11, 12, and 13.  The measured colors were used to estimate 
        absolute $K$ magnitudes based on the 12 color-magnitude relations 
        presented by \citet{Henry2004}.  The photometric distances 
        presented in Column 16 of Table \ref{tab:photometry} were derived 
        from the mean of the distance moduli implied by the absolute 
        $K$ magnitudes and the 2MASS apparent $K$ magnitude. 

        Relative photometry (for variability studies) comes from our parallax
        pipeline. With multiple nights of data in the filter used for
        parallax, we use the methods in \citet{Honeycutt1992} to derive the
        nightly offsets and zero points for relative instrumental photometry
        \citep{Jao2008} to derive stellar variability.  These values are given 
        in Column 8 of Table \ref{tab:photometry}.

    \subsection{Astrometry}
    \label{sec:astrometry}
        CTIOPI astrometry is carried out using the same telescope
        and camera configuration as that used for photometry
        (\S\ref{sec:photometry}) but uses only one filter for 
        each object, chosen to provide the best balance between 
        target(s) and reference star signal-to-noise ratio 
        values.  The astrometric pipeline uses all available 
        images taken at hour angles less than two hours, and 
        produces parallaxes, proper motions, and time-series 
        photometry in the parallax filter, all relative to 
        between 5 and 15 ``reference'' stars within a few 
        arcminutes of the target stars and visible in our 
        images.  Parallaxes were corrected to absolute values 
        (Columns 11 and 12 of Table \ref{tab:astrometry}) using 
        the mean of the photometric distances to the reference
        stars, with a typical correction of 1.5$\pm$0.5 mas.  For a
        small number of targets with seemingly nearby reference
        fields (mean photometric parallax esimate, $>3.0$ mas), we assume the 
        reference stars are actually reddened by some galactic source, and 
        instead apply the typical correction stated above.
        Between 2005 Mar and 2009 Sep, a different $V$-band filter 
        was used for astrometric and photometric observations.  While 
        photometrically identical to the original $V$ filter, it exhibited 
        slightly inferior astrometric performance \citep{Riedel2010}, 
        and all astrometric solutions that incorporate 
        data from it are marked as such in Column 16 of Table \ref{tab:astrometry}. 
        Additional details of CTIOPI observing procedures can be found in
        \citet{Jao2005}, \citet{Henry2006}, and other papers in this series.

   \subsection{Interferometry}
   \label{sec:interferometry}
        Four of the objects in this paper -- 
        BD-21$^{\circ}$1074BC, SCR~0613-2742AB, L~449-1AB, and SCR~2010-2801AB -- 
        were selected for their X-ray brightness and observed with the Hubble 
        Space Telescope's Fine Guidance Sensors in Cycle 16B, in proposal 
        11943/11944 (``Binaries at the 
        Extremes of the H-R Diagram'') using the F583W filter\footnote{The 
          bandpass of the F583W filter is shown here: 
          http://www.stsci.edu/hst/fgs/design/filters (checked 
          2013 Jun 04)} with no pupil.
        Reductions were carried out for both axes, providing sub-milliarcsecond 
        precision separations, and delta magnitudes (hereafter $\Delta mag$).  
        All four targets were found to be binaries (see $\S$ \ref{sec:systemnotes} 
        below) and their separations, position angles, and magnitude differences 
        were determined by fitting with single-star fringe scans as described 
        by \citet{Nelan2004}.

        Hubble's Fine Guidance Sensors (FGS) are implemented as three
        movable units equipped with Koesters prisms, which allow them
        to function as a 2D interferometer.  Two FGS units lock on
        guide stars and stabilize the spacecraft, while the
        third (FGS 1r) scans back and forth across the star.  Rather than 
        a Michelson interferometer, light is channelled through a linear
        polarizing beamsplitter, and then through the unit's Koesters
        prisms, where the light interferes with itself, allowing two
        dimensions of fringes to be read out simultaneously.

        As an interferometric instrument, FGS is capable of measuring
        single-axis separations as small as 8 mas, for objects as
        faint as $V$=16.8. It has been routinely used for orbital mapping and
        sub-milliarcsecond parallaxes \citep[e.g.][]{Franz1998,Benedict1999,McArthur2011}.

    \subsection{Spectroscopy}
    \label{sec:spectroscopy}
	\subsubsection{CTIO 1.5m RCSpec}
        Spectroscopic observations of all the resolved objects in 
        this paper (except SCR 0613-2742AB) were carried out 
        between 2003--2006 and 2009--2011 using the CTIO 1.5m telescope under 
        the aegis of the SMARTS Consortium.  The CTIO 1.5m 
        Richey-Chretien Spectrograph (RCSpec) was used with the 32/I grating setting,
        covering 6000--9600\AA~at a resolution of 8\AA.  Two spectra of each target 
        were taken back-to-back to allow cosmic ray rejection.
        Data were reduced with standard IRAF techniques using one flux
        standard per night for absolute flux calibration.  No telluric standards
        were observed, as the data were not originally intended for any
        purpose other than determining spectral types.

        Spectral types were determined by direct comparison to previously obtained
        standard stars \citep{Henry2002} spectra.  The spectra were
        prepared by interpolating them onto a fixed 1\AA~grid running 
        from 6000--9000\AA.  We then removed telluric features
        (defined as any region with $>4\%$ absorption from the
        \citet{Hinkle2003} sky absorption atlas) and H$\alpha$.  The
        target spectra were then compared to the standards by dividing 
        target by standard after cropping both spectra to only the wavelengths 
        where both spectra overlap.  The lowest standard deviation $stddev\left( 
        \frac{target}{standard} \right)$ was taken as the 
        best-matching spectral type.  The resulting spectral types are 
        presented in Column 14 of Table \ref{tab:photometry}, and have 
        $\pm$0.5 type errors. Offsets compared to other spectral typing methods 
        \citep{Reid1995} are of similar size.

        Spectral line equivalent widths and indices were computed with the
        same program, utilizing 11\AA~windows centered on the maximum
        or minimum of the feature for both H$\alpha$ and \ion{K}{1} 7699\AA.
        Full bins of 24\AA~were used for the \ion{Na}{1} doublet index.  These 
        features were measured prior to removing the telluric and H$\alpha$ 
        features.

        \subsubsection{CTIO 4.0m RCSpec}
        We obtained spectra of SCR~0613-2742AB with the CTIO 4.0m telescope's
        RCSpec on 2008 Sep 18 and 2008 Sep 19 using the KPGLF-1 (632
        g/mm) grating, which covers 4900--8050\AA~at a resolution of
        1.9\AA~per pixel.  The spectral type, H$\alpha$, and K {\sc
          I} EW measurements of SCR~0613-2742AB come from spectra
        taken at this telescope, and were calibrated using the same 
        program that analyzed the above CTIO 1.5m RCSpec data.

        \subsubsection{MPG 2.2m FEROS}
        One spectrum of SCR~0613-2742AB was taken with the FEROS 
        spectrograph \citep{Kaufer1999} on the MPG 2.2m telescope at La Silla Observatory 
        on 2013 Feb 18 as part of ESO program 090.C-0200(A). FEROS is an echelle 
        spectrograph fed by two 2.0\arcsec~fibers and provides 
        R$\sim$48,000 spectra over a wavelength range of 
        3500--9200\AA. Observations were taken in the Object-Sky mode 
        with the use of the atmospheric dispersion corrector. The 
        data were reduced with the facility pipeline and the IRAF 
        task {\it fxcor} was used to cross-correlate the target
        spectrum with several radial velocity standards observed in
        the same fashion. We measure a heliocentric radial velocity of
        $+$22.54$\pm$1.16 km s$^{-1}$ for SCR 0613-2742AB (Table 
        \ref{tab:kinematicstable}).  The \ion{Na}{1} gravity index measurement 
        and Li 6708\AA~EW for SCR~0613-2742AB were also derived from 
        these spectra.

        \subsubsection{CFHT ESPaDOnS}
        SCR~1425-4113AB was observed on the Canada-France-Hawaii Telescope (CFHT) with the ESPaDOnS 
        \citep{Donati2006}. ESPaDONs was used in the ``star+sky'' mode, 
        to get a resolving power of R=68000 covering 3700--10500\AA~over 
        40 grating orders. The data were reduced by the queue service 
        observing team using UPENA pipeline. We measure a heliocentric radial 
        velocity, lithium EW, and $v\sin i$ for both targets in the system.

\section{An Assessment of Stellar Youth Tracers}
\label{sec:analysis}


Our available data -- trigonometric parallaxes, low resolution red optical 
spectroscopy, and $VRIJHK$ photometry for the entire sample -- provides 
four methods of young star identification at our disposal: UVWXYZ 
Kinematics, Isochrones, the Bayesian Analysis for Nearby Young AssociatioNs 
(BANYAN) statistical method, and gravity sensitive features.  Taken 
individually, none offers conclusive proof of youth.  Together, they present 
a strong case for the youth (and specific association membership) of the 
systems described here.  The measured data used to make our conclusions are 
given in Table \ref{tab:kinematicstable}, Table \ref{tab:isochronetable}, 
and Table \ref{tab:youngproperties}; the results of our analysis are given 
in Table \ref{tab:youngresults}. 

    \subsection{UVWXYZ Kinematics}
    \label{sec:kinematics}
    Stars that form together should be moving together through space.  
    Over time, internal and external interactions will cause them to 
    disperse into the thin disk of the Galaxy.  The associations
    we consider here (Table \ref{tab:associationstable}) are all 
    sufficiently young that this has not happened yet, even though 
    (with the exception of $\eta$ Chameleontis, the Pleiades, and the 
    Hyades) they are gravitationally unbound.

    The perhaps hundreds of systems in an association are spread out 
    over tens of thousands of cubic parsecs, interspersed among 
    thousands of field systems (and, indeed, members of other young 
    moving groups), and as an unbound association, their velocity 
    dispersions are larger than 1 km s$^{-1}$.  Therefore, large numbers 
    of older stars (\citealt{Lopez-Santiago2009} in particular quotes 
    $\sim 30\%$) will coincidentally happen to have matching UVW motions, 
    and even larger numbers of field stars will fall within the spatial 
    boundaries of an association. UVW motions do not prove youth, but 
    they are necessary to connect young stars to young associations.

    There is also the non-trivial issue of whether we have an accurate 
    assemblage of these nearby associations and moving groups.  
    We adopt the associations in Table \ref{tab:associationstable} despite 
    the knowledge that the physical reality of these groups (and the 
    accuracy of their proposed members) is still somewhat uncertain.  
    For instance, the IC~2391 Supercluster \citep{Eggen1991}, 
    Carina-Vela moving group \citep{Makarov2000}, and Argus association 
    \citep{Torres2008} have all been proposed as the extended halo of the 
    nearby IC~2391 open cluster, but all are more or less distinct from 
    each other in terms of membership and proposed properties; currently only 
    Argus is thought to be an actual co-eval assembly.

    \subsubsection{The Kinematic Data}
    Computing UVWXYZ kinematics requires requires R.A., Decl., $\pi$, 
    $\mu_{R.A.\cos{Decl.}}$, $\mu_{Decl.}$, and radial velocities for UVW velocities; 
    and R.A., Decl., and $\pi$ for XYZ positions. The input data used for this 
    analysis (presented in Table \ref{tab:kinematicstable}) differs in several
    key ways from the pure CTIOPI astrometric data in Table \ref{tab:astrometry}:

    For most star systems, the only available spectra are from the 
    CTIO 1.5m RCSpec, which lacks the resolution necessary for radial 
    velocities.  We have obtained radial velocities from the literature to fill out our sample.  

    We use weighted mean parallaxes for systems that have 
    multiple reported parallaxes -- either from multiple system 
    components, or parallax determinations from other parallax programs.  
    In doing so, we have made two key assumptions: That all components of 
    a star system are at the same effective distance to currently achievable 
    accuracy; and that all published parallaxes are reasonably free of
    systematics.  

    The CTIOPI pipeline produces relative proper motions (as shown in Table 
    \ref{tab:astrometry}) by assuming each target field's astrometric reference 
    stars have no net motion, which is not strictly true.  In order to have unbiased 
    results, we ideally want to use absolute proper motions for our kinematic 
    determinations.  Comparing our results to the absolute International 
    Celestial Reference System (ICRS) proper motions\footnote{Note that UCAC4 also 
      includes high proper motion objects from relative proper motion sources, 
      including previous CTIOPI papers.  Such entries are identifiable by flags within 
      the database.} 
    from UCAC4 \citep{Zacharias2013}, we find mean offsets of 
    $+5.12\pm12.80$ mas yr$^{-1}$ in $\mu_{R.A.\cos{Decl.}}$ and $-0.78\pm9.45$ mas yr$^{-1}$ 
    in $\mu_{Decl.}$, uncorrelated across the sky.  The small differences but large 
    uncertainties suggest CTIOPI proper motions are accurate relative to the UCAC4 ICRS grid, but 
    the uncertainties are undersampled. Where possible, we use UCAC4 
    absolute proper motions directly. Where no UCAC4 proper motions 
    are available, we use our proper motions, with the above offsets 
    added in as a systematic uncertainty.  In two cases -- L~449-1AB and AP~Col -- 
    the UCAC4 proper motion was discrepant with the CTIOPI proper 
    motion by more than 100 mas yr$^{-1}$, and we used the CTIOPI 
    proper motion with the uncertainty.

    \subsubsection{The Kinematic Method}

    The standard method for computing UVW space velocities is laid out in 
    \citet{Johnson1987}, and the matricies in that paper can easily be adapted to compute 
    XYZ space positions.  These UVWXYZ coordinates are right-handed Cartesian Galactic coordinates aligned so that the U/X axis is toward the galactic center, the V/Y axis is in the direction of galactic rotation, and the W/Z axis is toward the North Galactic Pole.  In cases where we have full kinematic information, we calculate $10^7$ Monte Carlo iterations to fully sample the uncertainties on our input kinematics as a three-dimensional ellipsoid in velocity space.  For stars without radial velocity measurements, we calculate $10^5$ Monte Carlo iterations at multiple different radial velocities within a range $-$100 to $+$100 km s$^{-1}$.

	To determine whether a star system is a potential match for a given association, we must determine how close the three-dimensional velocity-space ellipsoid(s) defined by our Monte Carlo iterations is to the three-dimensional velocity-space ellipsoid of the association, as defined in Table \ref{tab:associationstable}.  Because the dispersions are meaningful --  in the case of the sample system, they represent the uncertainty on the velocity of the system; in the case of the association, they represent the intrinsic dispersion of real members -- our phase-space ``separations'' are calculated relative to those dispersions, in the form of our goodness-of-fit statistic $\gamma$:
\begin{equation}
\gamma = \frac{1}{3} ( \frac{(U_{assoc}-U_{system})^{2}}{(\sigma_{U_{assoc}}^{2} + \sigma_{U_{system}}^{2})} + \frac{(V_{assoc}-V_{system})^{2}}{(\sigma_{V_{assoc}}^{2} + \sigma_{V_{system}}^{2})} + \frac{(W_{assoc}-W_{system})^{2}}{(\sigma_{W_{assoc}}^{2} + \sigma_{W_{system}}^{2})} )
\end{equation}
This $\gamma$ statistic is effectively identical to the one used in \citet{Shkolnik2012}, where it appears as $\bar{\chi}^2$. 

We take a value of $\gamma$ less than 4 as a 
    potentially significant match.  In many cases, a system will be 
    kinematically consistent with membership in more than one 
    young association.  This is often unavoidable, as the velocity 
    distributions of several associations genuinely overlap; in
    these cases we must look at the other diagnostics to determine which 
    association is the most consistent with the available data.
In cases where we have no radial velocity, we fit the resulting $\gamma$ values from our range of points to determine a best-fit radial velocity and $\gamma$.
    Only the $\gamma$ value for the most consistent association is given 
    in Column 5 of Table \ref{tab:youngresults}.

    \subsection{Isochrones}
    \label{sec:isochrones}
    Pre-main-sequence stars are still in the process of contracting
    under gravity, and are still physically larger than main-sequence
    stars of the same temperature, and consequently much brighter.  It
    is therefore possible to distinguish stars of a given age using
    isochrones, or at least demonstrate the youth of a system.  

        \subsubsection{The Photometric Data}
        \label{sec:deblending}
        Overluminosity is also a sign of multiplicity, and must therefore be 
        taken into account before making conclusions about the potential 
        youth of a system based on its luminosity.
        Our sample has many multiple stars (See Table \ref{tab:multiples}),
        which are not resolved in CTIOPI images if they are less than
        $\sim$2\arcsec~apart.  In order to properly distinguish between binaries 
        (which in the maximal case of an equal-luminosity pair can be 0.7 
        mag brighter than a single star) and young stars, we have 
        made an extensive literature search for multiples within the sample.  
        Once multiples are identified, their photometry must be de-blended 
        to properly place them on a color-magnitude diagram.

\begin{figure}
\center
\includegraphics[angle=90,width=.4\textwidth]{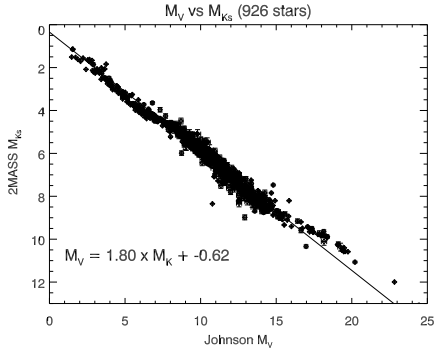}
\caption{The approximately linear relationship between Johnson $M_V$ and 2MASS $M_{K_S}$, as 
determined from 962 stars with trigonometric parallax errors less than 10 mas. The values here 
can be used to convert $\Delta$ V measurements into $\Delta K_s$ and vice versa. 
$V$ data from \citet{Bessell1990a,Bessell1991}, and previous papers in this series; $K_S$ 
data from 2MASS \citep{Cutri2003}.  Other relations are given in Table \ref{tab:deblendingtable}.
\label{fig:deblending}}
\end{figure}

        Unfortunately, we do not have both $\Delta V$ and $\Delta K_S$ values 
        for any multiples in our sample, so the deblending in one of the 
        filters must be estimated.  Plotting, for instance, $M_V$ vs 
        $M_{K_{S}}$ (Figure \ref{fig:deblending}) demonstrates that the 
        relation is, to a first approximation, linear between $M_V$=1 and 
        $M_V$=15 ($M_{K_S}$=1 and $M_{K_S}$=9), in that case with a slope of 
        1.8, for $\Delta V = 1.8 \times \Delta {K_S}$.   We list all the 
        equivalencies in Table
        \ref{tab:deblendingtable}.  Actual measurements should
        definitely be preferred, however, as the error on the fit
        slopes are on the order of 0.03, and the residuals
        to the fits are on the order of 0.5.

        For purposes of deblending our stars, we have assumed that the $\Delta FGS 583W$ 
        values are equivalent to Kron-Cousins $\Delta R$, and 
        that all the various $\Delta K$ values (CIT, Altair, MKO, 2MASS) are 
        equivalent to 2MASS $\Delta K_{S}$. Several systems are known only 
        as spectroscopic binaries.
        With no other information available, we have assumed $\Delta V = \Delta K_{S}$ =0.
        The photometric data used in the isochrone analysis, including deblended 
        magnitudes, are given in Table \ref{tab:isochronetable} and 
        used to place stars on the Color-magnitude diagram in Figure 
        \ref{fig:isochrones}.  Deblended optical magnitudes are given for 
        the brown dwarfs TWA 27B and SCR 0103-5515C, but their deblended 
        $V$ magnitudes are suspect at best.

        \subsubsection{The Photometric Method} 
        Because isochrones do not match field main sequence stars well at low 
        masses \citep{Hillenbrand2004}, we are using empirical ``isochrones'' 
        from \citet{Riedel2011}, which are empirical polynomial fits derived 
        from known young stars with photometry and trigonometric parallaxes, 
        for the four associations -- TW Hya, $\beta$ Pic, Tuc-Hor, and AB Dor -- 
        with a sufficient number of known low-mass members to make the 
        isochrones useful. Plotting our stars on color-magnitude diagrams 
        with these isochrones (Figure \ref{fig:isochrones}) demonstrates that 
        many are indeed overluminous relative to main sequence stars.  Note that 
        several stars (SCR~0103-5515C, LP~655-48, TWA~27A and B) are too red and 
        faint to appear in the figures, and the Tuc-Hor isochrone line from 
        \citet{Riedel2011} does not extend into the M dwarfs shown due to lack of data.

    \subsection{BANYAN}
    \label{sec:banyan}
    BANYAN \citep{Malo2013} is an independent Bayesian methodology for finding 
    young stars.  BANYAN uses $IJ$ photometry, \citet{Baraffe1998,Baraffe2002} 
    model isochrones, and a slightly different set of UVWXYZ values for the 
    known associations.  It searches for members of the known nearby 
    associations $\beta$ Pictoris, TW Hydra, Tucana-Horologium, Columba, 
    Carina, Argus, and AB Doradus, with ``Field'' as the default hypothesis.
    It is presented as a complementary, proven method.  The coefficients 
    for the best-matching association are shown in Table 
    \ref{tab:youngresults}, Columns 5 and 6.

    \subsection{Low Surface Gravity}
    \label{sec:gravity}
    Three gravity-sensitive features exist within the 6000\AA--9000\AA~coverage 
    of our spectra: Ca~II (8498,8542,8662\AA) is strong in giants and 
    weak in dwarfs; \ion{Na}{1} (8183,8195\AA) and \ion{K}{1} 
    (7665,7699\AA) are weak in giants and strong in dwarfs.
    The general pattern is that the neutral alkali and alkali earth metals
    are increasingly strong with higher gravity, while their singly
    ionized variants grow weaker with higher gravity \citep{Allers2007,Schlieder2012b}.

\begin{figure}
\center
\includegraphics[angle=0,width=.6\textwidth]{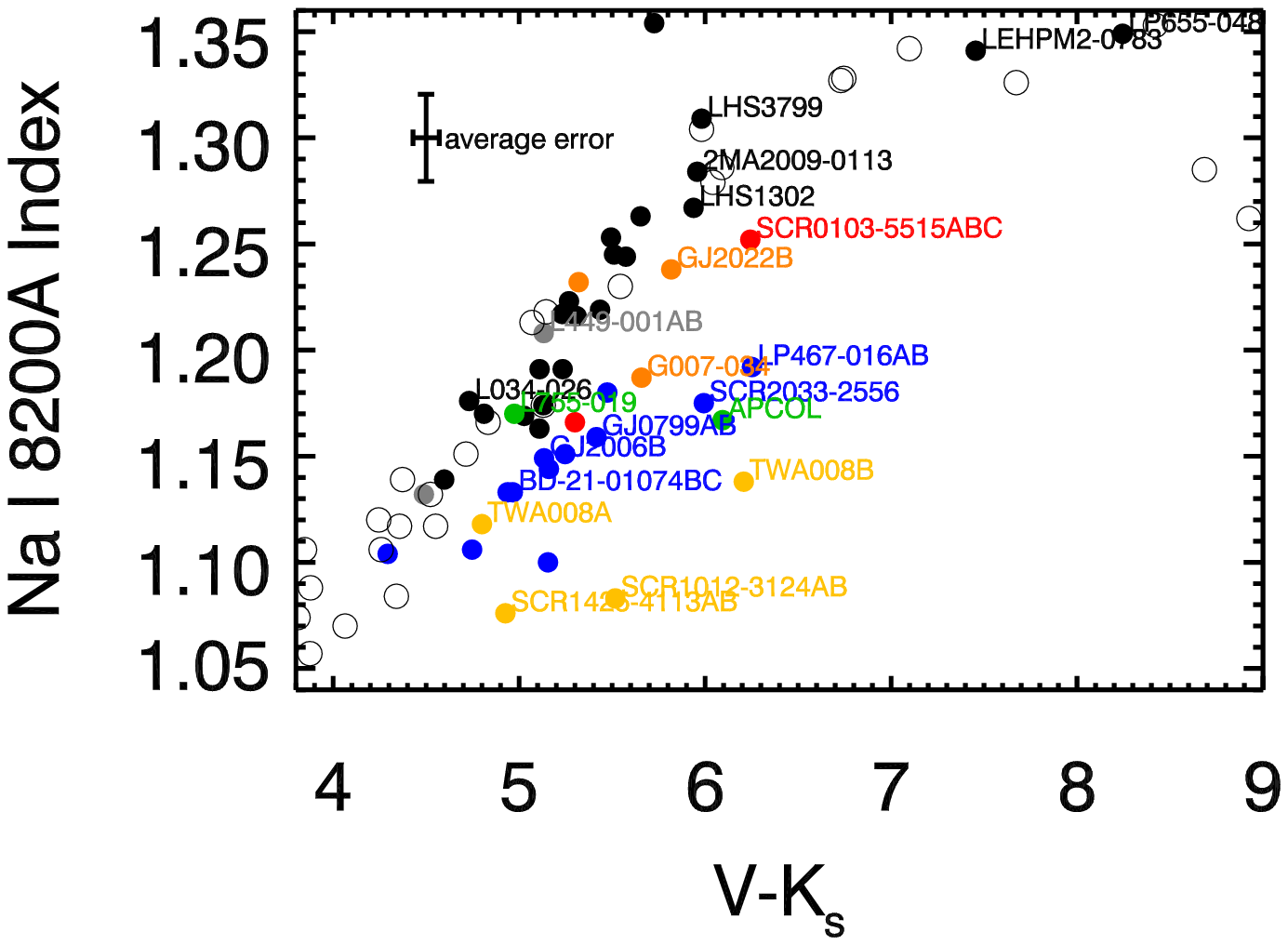}
\includegraphics[angle=0,width=.6\textwidth]{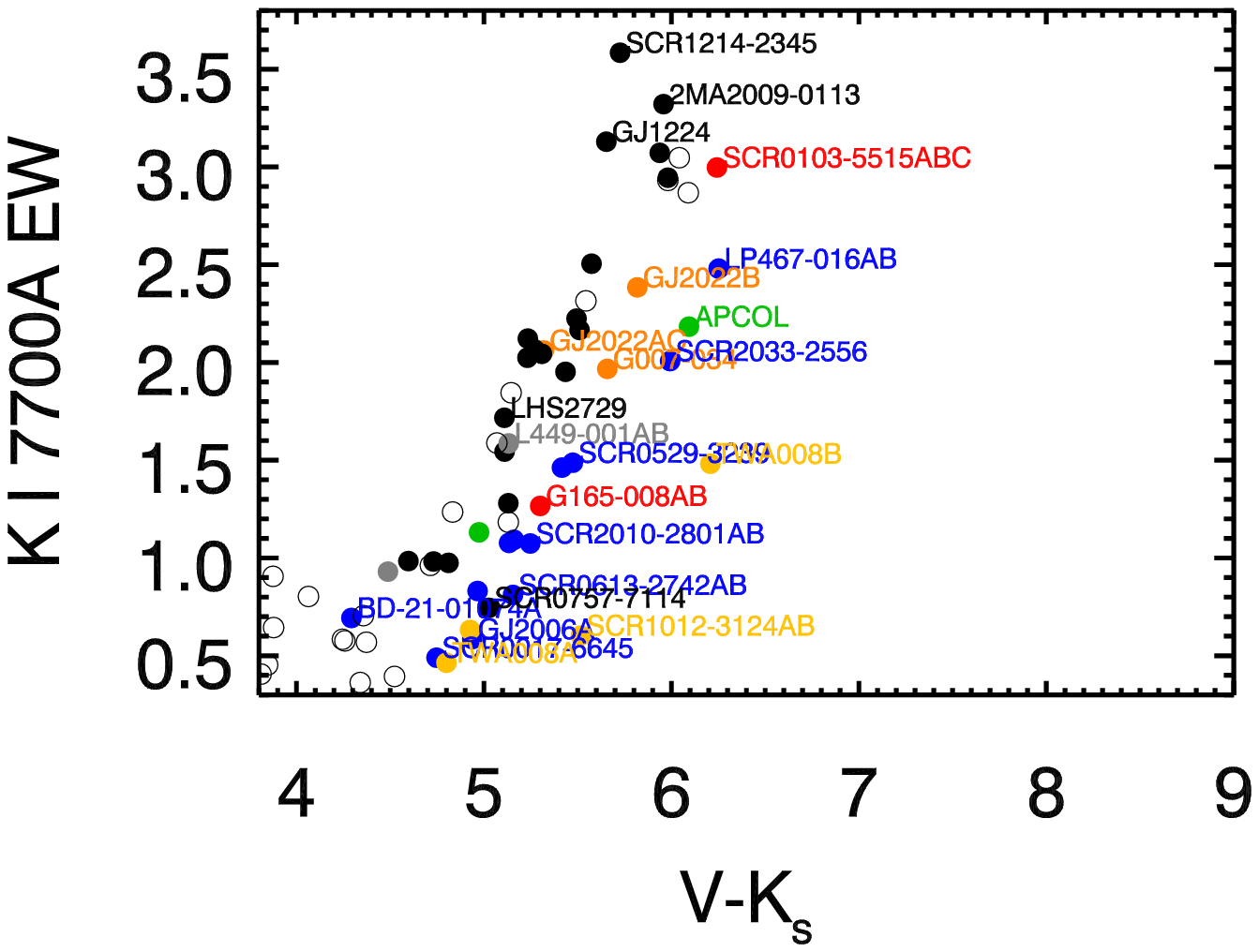}
\caption{The Na {\sc I} 8183/8195\AA~doublet gravity index from
  \citet{Lyo2004a} (above) and K {\sc I} 7699\AA~EW (below) vs $V-K_S$ 
  color.  Open circles are known main-sequence stars from other RECONS 
  spectroscopic efforts, for comparison. The stars are colored by the 
  association (if any) they appear to belong to: TW Hya (Yellow), $\beta$ 
  Pic (Blue), Tuc-Hor/Columba/Carina (Red), Argus (Green), AB Dor (Orange), 
  Castor/UMa/Hya (Gray), Field (Black).
  \label{fig:gravity}}
\end{figure}

    In this paper we are using the \ion{Na}{1} index as defined in
    \citet{Lyo2004}, and the \ion{K}{1} 7699\AA~doublet line (the other is 
    contaminated by the atmospheric A band).  The \ion{Na}{1} index value 
    is formed by the ratio of the average flux in two 24 \AA~wide bands:
    \begin{equation}
      Na I_{index} = \frac{F_{8148-8172}}{F_{8176-8200}}
    \end{equation}
    Measurements for the program stars are given in Table 
    \ref{tab:youngproperties}, and graphs of the general trends are 
    shown in Figure \ref{fig:gravity}. Unfortunately, as can be seen in
    the figures, young stars and main-sequence stars overlap at colors 
    bluer than $V-K_s=5$, in both cases.  The lines can also be affected by
    stellar activity, where emission fills in the absorption line cores,
    leading to lower EWs \citep{Reid1999}.

    In principle, multiplicity and metallicity will have an effect on
    these features.  The effects are muted in the case of
    multiplicity, as the brighter component will dominate, and the line 
    will not be appreciably weaker or stronger than that of a 
    main-sequence star of the same color.  Metallicity is more difficult.
    While we can expect low-metallicity subdwarfs to have weaker lines 
    due to lower abundances, subdwarfs plot {\it below} the main sequence 
    and will not be mistaken for pre-main-sequence stars.  However, as 
    noted by \citet{Shkolnik2009}, high metallicity stars of a given 
    mass and bolometric luminosity will masquerade as stars of lower
    temperature {\it and} lower surface gravity. The additional metals
    increase the opacity of the stellar atmosphere and therefore put the
    effective photosphere farther from the center of the star. Thus, these 
    stars will appear above the main sequence -- in fact, these 
    gravity-sensitive atomic species are the ones used by 
    \citet{Rojas-Ayala2010} to measure the metallicity of field M dwarfs.

    With the exception of SCR~0613-2742AB and SCR~1425-4113AB, the only 
    available spectra for the sample stars 
    are low resolution flux-calibrated optical spectra from the
    CTIO 1.5m telescope that cannot be reliably corrected for telluric
    absorption.  Thus, there is an additional source of error in our
    \ion{Na}{1} index measurements, and we are using only one of the \ion{K}{1}
    doublet lines (the other is within the atmospheric A band).  We conclude 
    that while we see some indication of offsets for young stars in the plots 
    of Figure \ref{fig:gravity}, conclusions of youth using this method with 
    data at this resolution are only tentative.

    \subsection{Activity-based features}

    \subsubsection{X-ray activity}
    \label{sec:xrays}
    As seen in \citet{Zuckerman2004}, X-ray activity remains saturated 
    in M dwarfs even beyond 600 Myr. Consequently, X-ray activity is 
    a considered as a necessary but insufficient marker of a star system's 
    youth, and its presence was mostly useful in our sample selection process.

    There are, however, eight objects in our sample without X-ray 
    emission.  These objects comprise the six previously mentioned 
    systems with no X-ray emission, and two components (GJ~2022B and 
    LP993-115A) of systems with other X-ray detected components.  For 
    those targets, we calculate upper limits on their X-ray emission (see Table 
    \ref{tab:youngproperties}, Columns 4 and 5) using the cnts s$^{-1}$ arcmin$^{2}$ for 
    the nearest target in the RASS catalog (generally under 
    2\arcmin~distant) as the local background countrate.  Even assuming 
    a hypothetical one-arcminute point spread function, of the eight 
    objects, only TWA~27AB can potentially have saturated X-ray emission, 
    unless the background count rates for ROSAT are remarkably different 
    over small angular separations.  For the special case of NLTT~372, 
    G~131-26AB is also within the 25\arcsec~radius we used to localize 
    X-ray detections.  Although the emission is more likely associated 
    with G~131-26AB, if the observed X-ray counts are actually 
    being produced by NLTT~372, it would be one of the most coronally active 
    stars in our sample.

    ROSAT's resolution and accuracy cannot spatially distinguish 
    between targets that are within $\sim
    25$\arcsec~of each other.  Consequently, there are a few cases where a
    multiple system is resolved in the optical and near-infrared but 
    not in X-rays. In these cases, we have {\it combined} the $V$ and 
    $K_S$ photometry to produce a system bolometric flux.  We did not blend 
    NLTT~372 and G~131-26AB, because despite being arcseconds from each 
    other, they are two separate star systems.

    \subsubsection{H$\alpha$ emission}
    \label{sec:Halpha}
\begin{figure} 
\center
\includegraphics[angle=0,scale=.6]{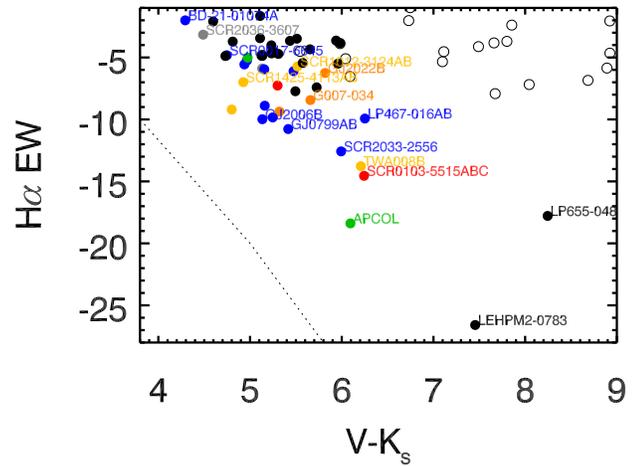}
\caption{H$\alpha$ EW versus $V-K_S$ color.  The dashed line is (roughly) 
the dividing line between Classical T Tauri stars, as defined in 
\citet{White2003}. Classical T Tauri stars would appear in the lower left, 
below the line; none of the stars (even known young stars like 
RX~1132-2651AB=TWA~8AB) are potential Classical T Tauri stars.
The stars are colored by the association (if any) they appear 
to belong to: TW Hya (Yellow), $\beta$ Pic (Blue), Tuc-Hor/Columba/Carina 
(Red), Argus (Green), AB Dor (Orange), Castor/UMa/Hya (Gray), Field 
(Black).  Open circles are main-sequence stars from other RECONS 
spectroscopic efforts.
  \label{fig:Halpha}}
\end{figure}

    M dwarfs also have saturated H$\alpha$ emission for long periods
    of time \citep{West2008}.  We have measured this
    value (Table \ref{tab:youngproperties}),
    but it cannot be used to distinguish between the ages of low-mass 
    pre-main-sequence stars.  The only useful purpose of H$\alpha$ 
    emission for low mass stars is to provide an upper (if present) 
    or lower (if absent) limit on ages, which has been done in done 
    in Table \ref{tab:youngresults} following the prescription in
    \citet{West2008}. Though we see 
    substantial H$\alpha$ emission in a few of our stars, none of their 
    emissions reach the \citet{White2003} threshold necessary to be 
    considered a T Tauri star (see Figure \ref{fig:Halpha}).

    \subsubsection{Photometric Variability and Flares}
    \label{sec:variability}
\begin{figure}
\centering
\includegraphics[angle=90,width=.5\textwidth]{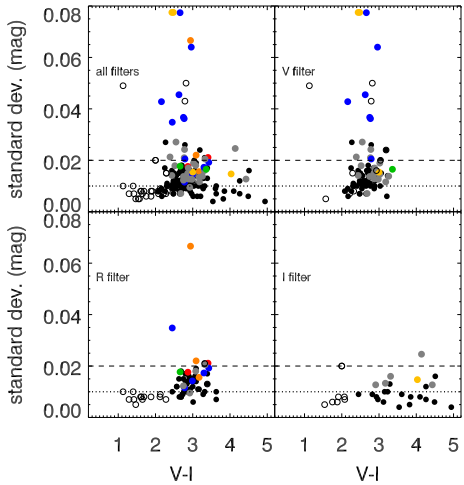}
\caption[Variability of Sample]{A plot of the relative
variability of the stars in this paper (closed symbols, colored as 
in previous figures), compared to previously published CTIOPI stars 
(black open symbols), as in \citet{Jao2011}.  Note the different scale 
from \citet{Jao2011}.  Our variability ``floor'' is 0.007 mag, so 0.020 
mag clearly indicates a variable star.  \label{fig:variability}}
\end{figure}

    One of the hallmarks of the T Tauri class of young stars
    is variability, and this extends into the older non-accreting
    stars discussed here.  Analysis of the relative variability of 
    M dwarfs \citep{Jao2011} shows that a typical M dwarf varies by 
    roughly 0.010 magnitudes, and statistically significantly more 
    in $V$ and $R$ (0.013 mag) than $I$ (0.008 mag). They also 
    found a statistically significant difference between regular M 
    dwarfs and their older, metal poor subdwarf cousins (variability 
    0.007 mag, our observational ``floor'') which points to some combination 
    of age and (possibly) 
    metallicity influencing the amplitude of stellar variability.  As 
    seen in Figure \ref{fig:variability}, M dwarfs with variability 
    higher than 0.020 mag are rare, although many are present among 
    the young stars in this paper.

\begin{figure} 
\centering
\includegraphics[angle=90,width=.5\textwidth]{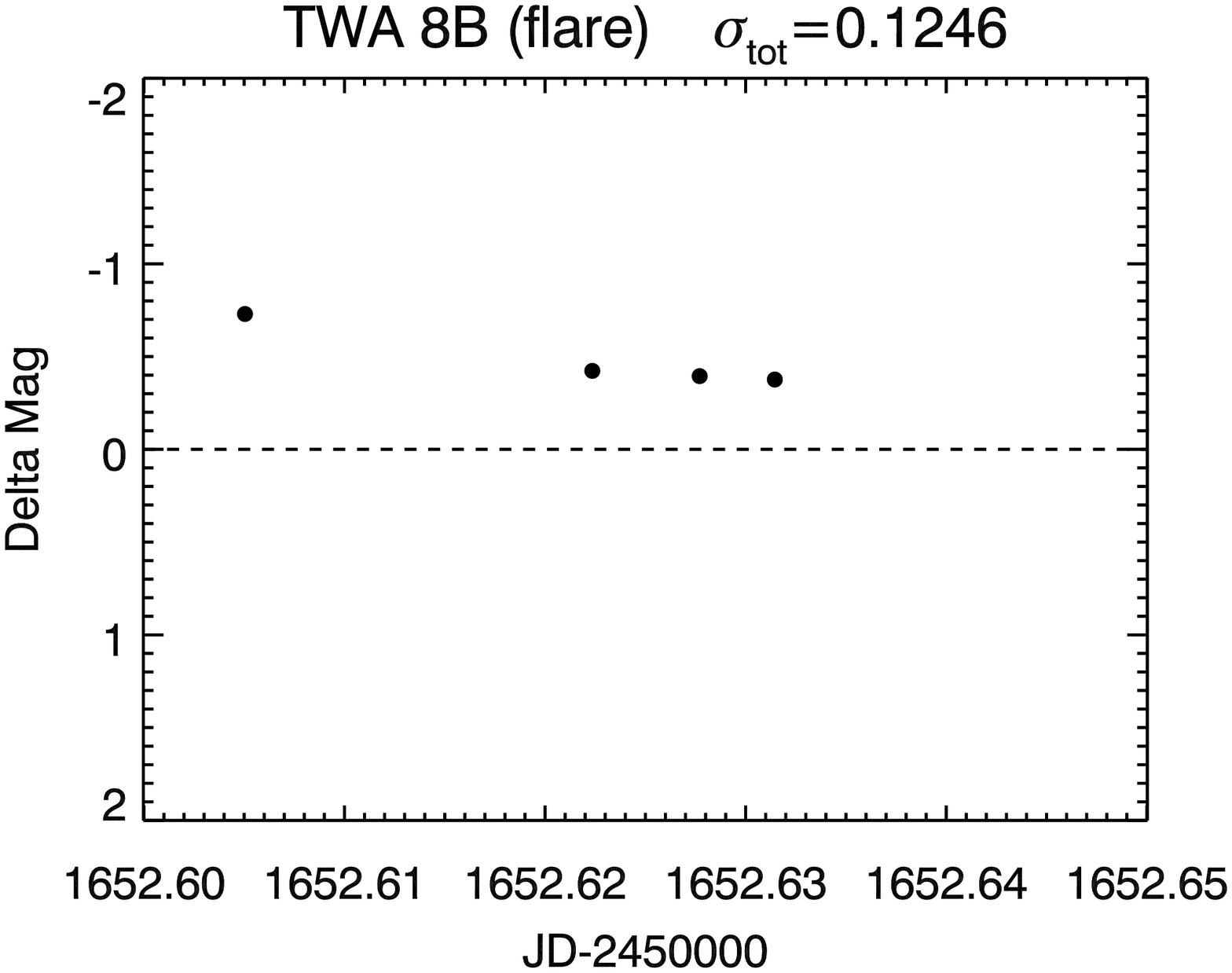}
\includegraphics[angle=90,width=.5\textwidth]{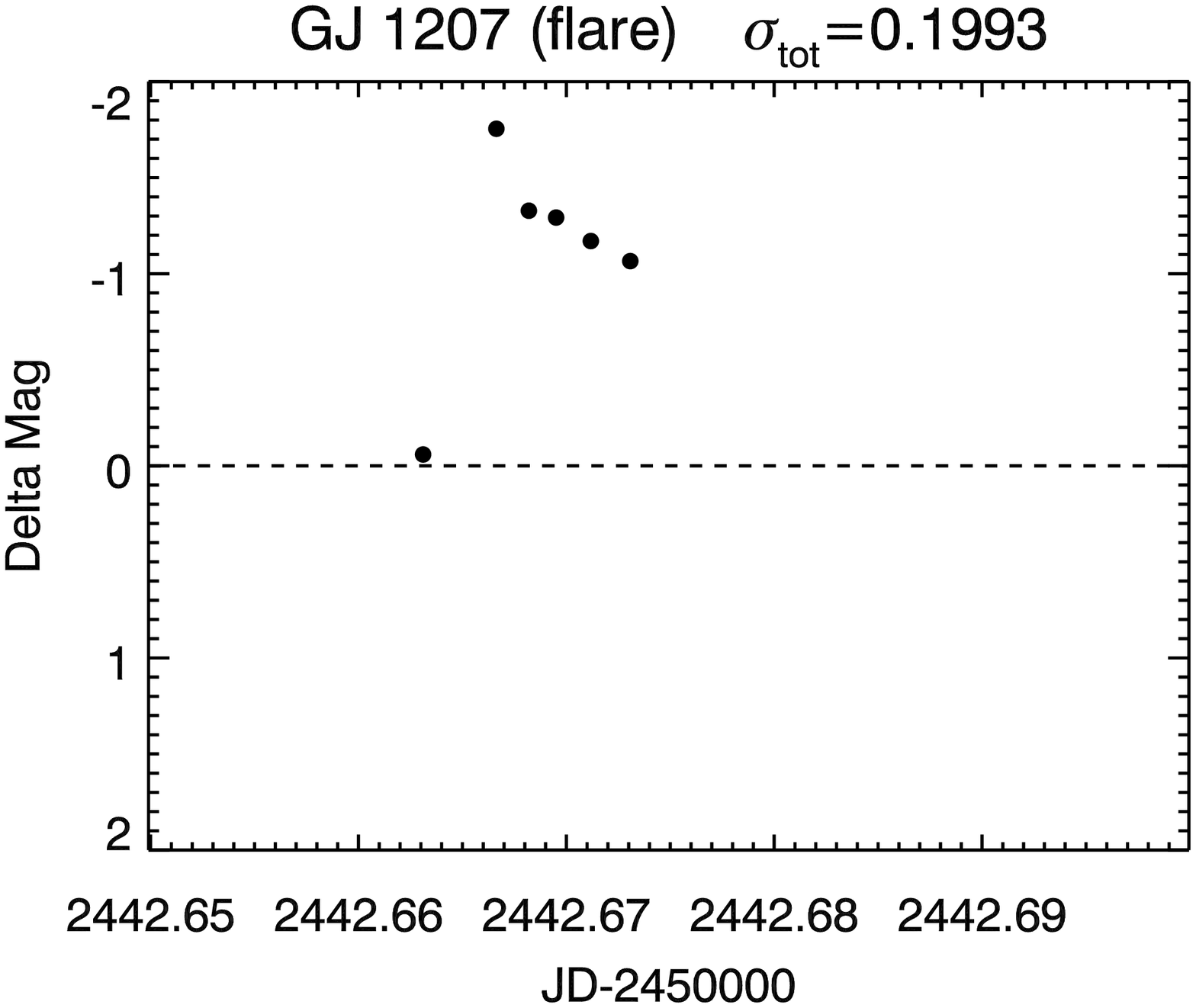}
\caption[Flares]{Two large partial flares observed in CTIOPI; Note the
timescales of the observations and the magnitude spreads; each was 
observed for less than an
hour.  We observed the peak of the GJ~1207 flare from 2002 Jun 17
(which rose in less than 5 minutes), but probably not the peak of the TWA 8B flare from
2000 Apr 18.  Given that each was observed for less than an hour, it
is difficult to tell which had a higher peak, was longer-lasting, or
had more total energy.  The $\sigma_{tot}$ scatter values given in Table 
\ref{tab:photometry} and Table \ref{tab:youngproperties} are highly
biased by these flare events.  The zero point of the delta magnitudes
were set by the other reference stars in the astrometric solution.
  \label{fig:flares}}
\end{figure}

    While many stars in this paper are known flare stars or UV Ceti
    variables, only two stars were seen to flare during the course 
    of astrometric observations: TWA 8B and GJ 1207.  The relative 
    photometry during the flares is reproduced in Figure \ref{fig:flares}.

\section{Candidate New Association Members}
\label{sec:results}

Of the 45 systems in this paper, 25 have their first trigonometric
parallaxes published here.  15 of the 45 systems (including 
LP~476-207=HIP~23418) are now new members of the sample of 
all stars with trigonometric parallax distances within 25 pc.

Seven systems are known triples (including GJ~799AB = AT Mic, whose 
primary is the star GJ~803 = AU Mic) and 13 are known 
binaries (Table \ref{tab:multiples}). 25 systems are currently not known 
to be multiple, though four (NLTT~372, SCR~0757-7114, 
LEHPM2-0783, SCR~2033-2556) are strongly suspected to be multiple based on our 
youth analysis, and one (SCR~0103-5515ABC) seems to be too bright 
even when its multiplicity is taken into account.  Thus, these 45 systems
include at least 72 objects, for a companion fraction (companion 
stars/systems) of 60\% and a multiplicity fraction (multiple 
systems/single systems) of 47\%, similar to what was reported by
\citet{Fischer1992} for a collection of M0--M3 dwarfs.  This high 
multiplicity rate is to be expected, considering that luminosity (the 
result of youth or multiplicity) was a defining characteristic of the 
selection process.

We find 21 potential and/or confirmed members of nearby young 
associations (Table \ref{tab:youngresults}, Column 10). Within 
the sample, we recover ten members of $\beta$ Pictoris, four 
members of TW Hydra, two members of Argus, two members of AB Dor, and two young members of unknown 
associations.  In addition, there is one system kinematically 
consistent with the Ursa Majoris moving group, although by gravity 
and HR diagram position it is not distinguishable from a main 
sequence star.  Particularly noteworthy are the seven new members 
of $\beta$ Pictoris.  Several systems are potentially kinematically 
consistent with the Hyades, but as all are more than two tidal radii (10 
pc) from the Hyades, and the existence of the Hyades Stream has been 
brought into question by \citet{Famaey2008}, we suspect all potential 
Hyades emberships are not physical (see $\S$\ref{sec:systemnotes}).

Several of our objects are likely cooler than the threshold where M dwarfs 
become fully convective (around $V-K=5.8$), and at least three (SCR~0103-5515C, 
TWA~27A, TWA~27B) are brown dwarfs.  It is difficult to make comparative 
conclusions about such systems, as very  few comparably low-mass systems are 
known in these associations.  

For AB Dor, Ursa Major, and Castor, our analysis is more tentative,
given that members of those systems are difficult to distinguish from
main sequence stars using isochrones and gravity-sensitive lines,
leaving us with only the kinematic analysis.  Confirmation of
membership in these associations will require measurements of the
stars' radial velocities, lithium, chemical abundances 
\citep{Castro1999,Barenfeld2013}, and 
(in the case of Castor) further study to determine if the moving 
group genuinely exists at all \citep{Mamajek2013}.

\subsection{Notable Systems}
\label{sec:systemnotes}
(in order of Right Ascension)

{\bf NLTT~372} was originally part of the reference field of the
system G~131-26AB.  It is within the error circle
of the ROSAT X-ray source we attribute to G~131-26AB, and may
either contribute or be the source of those X-rays.  It is more
luminous than a single star of its photometric colors, and is likely
to be a binary.

{\bf GJ~2006AB} appear to be $\beta$ Pictoris members, based on 
kinematics, gravity, and their positions relative to the $\beta$ Pictoris 
isochrone.  A spectrum of GJ 2006A was obtained with VLT instrumentation 
(Malo et al. in preparation) and the radial velocity was determined to 
be $+$8.90 km s$^{-1}$ with $v\sin i$ of 6 km s$^{-1}$, in excellent 
agreement with that of a predicted $\beta$ Pictoris member. 

{\bf SCR~0103-5515ABC} was resolved as a close triple by \citet{Delorme2013} 
composed of two M dwarfs, A and B, and a more spatially distant brown dwarf, 
C (see Table \ref{tab:multiples}).  \citet{Delorme2013} and the BANYAN results 
find it to be a match to Tucana-Horologium, but we find that its kinematics 
better match the Carina association, and because both associations are 
supposed to be the same age, we cannot distinguish between them with the other methods.
When their photometry is deblended, both components lie {\it above} the $\beta$ Pictoris 
isochrone, implying that they are younger than $\beta$ Pic, even though 
Tucana-Horologium and Carina are both supposed to be older than $\beta$ Pic.  The system is
undoubtedly young, but more observations are needed to determine if it
is a higher-order multiple in Tuc-Hor, or something entirely different.

{\bf GJ~2022ABC} is a hierarchical triple (see
Table \ref{tab:multiples}) composed of a wide (37.8\arcsec)
companion (B) to a close (1.8\arcsec) nearly-equal-luminosity pair (AC)
with a delta magnitude of roughly $\Delta V$=0.08 \citep{Jao2003,Daemgen2007}.
The B component is actually the least luminous, though to preserve the
historical order of discovery, we continue to refer to it as `B'.  The 
AC component is bright in RASS; there is no corresponding detection of the B component.

\begin{figure}
\centering
\includegraphics[angle=0,width=.4\textwidth]{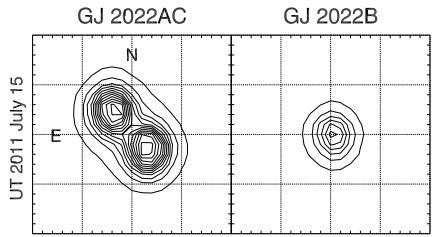}
\caption[GJ~2022AC]{Contour plot of GJ 2022A (South) and C (North) on
2011 Jul 15 from CTIOPI.  GJ 2022B (37.8\arcsec~distant) is also plotted as an
example single-star PSF, with the same contour intervals.  Grid lines
are 2.005\arcsec~apart (5 pixels at the CTIO 0.9m).
  \label{fig:GJ2022AC}}
\end{figure}

The A and C components are separated by 1.8\arcsec~(Figure
\ref{fig:GJ2022AC}). They are separable and (as they are much 
brighter than the B component) unsaturated on only 31 of the 66 
images taken.  Those 31 images were taken on 14 nights spanning 
11.89 years and positions were obtained using special 
extra-sensitive (but less precise)
SExtractor settings.   A reduction of all three components yields
\begin{itemize}
\item GJ~2022A: $\pi=$40.91$\pm$5.64 mas, $\mu=$210.5$\pm$1.3 mas
yr$^{-1}$ @ 126.5$\pm$0.71$^{\circ}$
\item GJ~2022B: $\pi=$42.12$\pm$3.60 mas, $\mu=$206.2$\pm$0.8 mas
yr$^{-1}$ @ 127.4$\pm$0.46$^{\circ}$
\item GJ~2022C: $\pi=$46.96$\pm$5.31 mas, $\mu=$200.6$\pm$1.2 mas
yr$^{-1}$ @ 126.8$\pm$0.71$^{\circ}$
\end{itemize} 

This implies a weighted mean parallax of 43.05$\pm$2.63 mas
(23.2$\pm$1.4 pc), which is consistent with our main reduction 
of B in Table \ref{tab:astrometry} using 66 frames (38.80$\pm$2.13 mas,
25.8$\pm$1.4 pc) and with membership in the AB 
Dor association.  Formally, we are using the
parallax of the B component alone as our system parallax in Table 
\ref{tab:kinematicstable} due to the lower precision of the results for 
all three components. 

\begin{figure}
\center
\includegraphics[angle=90,width=.5\textwidth]{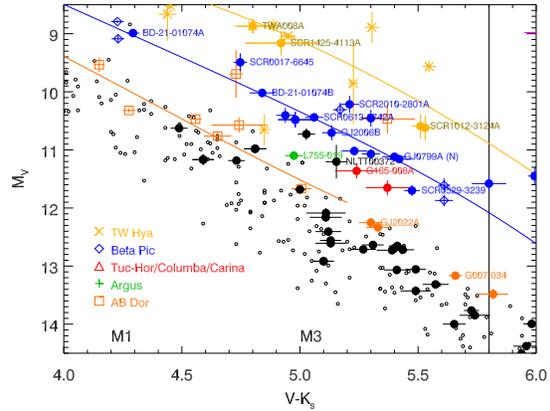}
\includegraphics[angle=90,width=.5\textwidth]{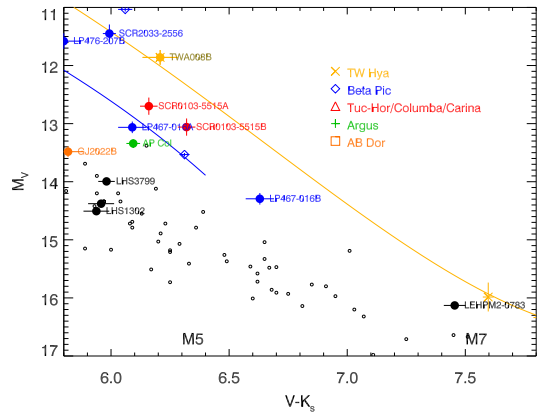}
\caption{The known components of the 45 systems described in this 
paper (excepting the extremely red objects TWA 27AB, LP 655-48, and 
SCR~0103-5515C), deblended where necessary and 
plotted relative to the RECONS 10 pc sample (open circles) on two 
overlapping color-magnitude diagrams.  Also
plotted are members of nearby young associations: TW~Hya (Xs), 
$\beta$~Pic (diamonds), Tuc-Hor (triangles), and AB~Dor (squares).  
Fifth order fits are plotted for (top to bottom) TW~Hya, $\beta$ Pic, 
and AB~Dor.  The line at $V-K_S$=5.8 is roughly the point at 
which M dwarfs are expected to become fully convective.
\label{fig:isochrones}}
\end{figure}

{\bf LHS 1302} is potentially a kinematic match for the $\sim$30 Myr old 
Columba association, but its gravity (Figure \ref{fig:gravity}) and 
isochrones (Figure \ref{fig:isochrones}) suggest it is a field object.

\begin{figure}
\centering
\includegraphics[angle=90,width=.4\textwidth]{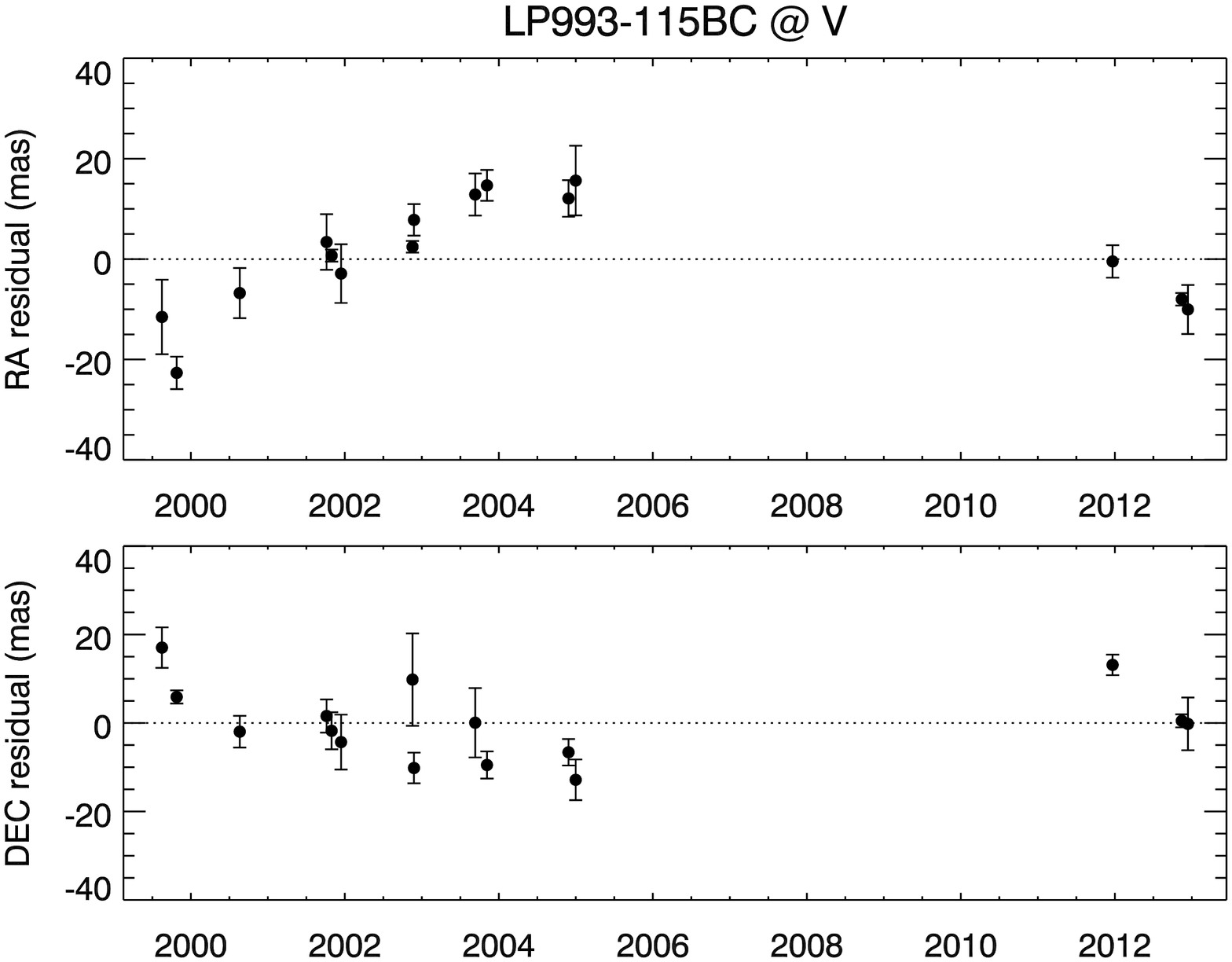}
\caption[LP~993-116AB astrometric perturbation]{Nightly mean residuals
of the LP~993-115BC (LP~993-116AB) parallax fit (after the parallax
and proper motion have been removed) show a clear astrometric
perturbation in both R.A. and Decl. axes.  This system is a known binary,
but the orbit has not wrapped, and our orbit fit does not converge.
  \label{fig:LP993-115BC_orbit}}
\end{figure}

{\bf LP~993-115 (A)/LP~993-116AB (BC)}
We detect the astrometric orbital motion of the BC pair (Figure
\ref{fig:LP993-115BC_orbit}).  Our attempt to fit an orbit did
not converge, and the astrometry in Table \ref{tab:astrometry}
was calculated without compensating for orbital motion.  This 
triple system appears to be composed of field stars, and only 
the BC components have X-ray emission.

{\bf LP~476-207ABC} was observed by {\it HIPPARCOS} as HIP~23418; the
resulting parallax was of poor quality due to erroneous coordinates in
the input catalog \citep{Perryman1997} and the resulting poor available 
astrometry.  Our new result (24.6$\pm$1.3 pc) 
is significantly closer and higher precision than the old \citep{Perryman1997} (32.1$\pm$8.8 pc) 
and revised \citep{van-Leeuwen2007} (33.2$\pm$10.5 pc) 
{\it HIPPARCOS} values, but the system is still in $\beta$ Pic.

{\bf BD-21$^{\circ}$1074ABC} is a known member of the $\beta$ Pic association
\citep{Torres2008,Malo2013}.  The A component was not originally targeted for 
parallax measurement and was saturated in many early images; the parallax 
result in Table \ref{tab:astrometry} is of lower precision.

\begin{figure}
\centering
\includegraphics[angle=90,width=.4\textwidth]{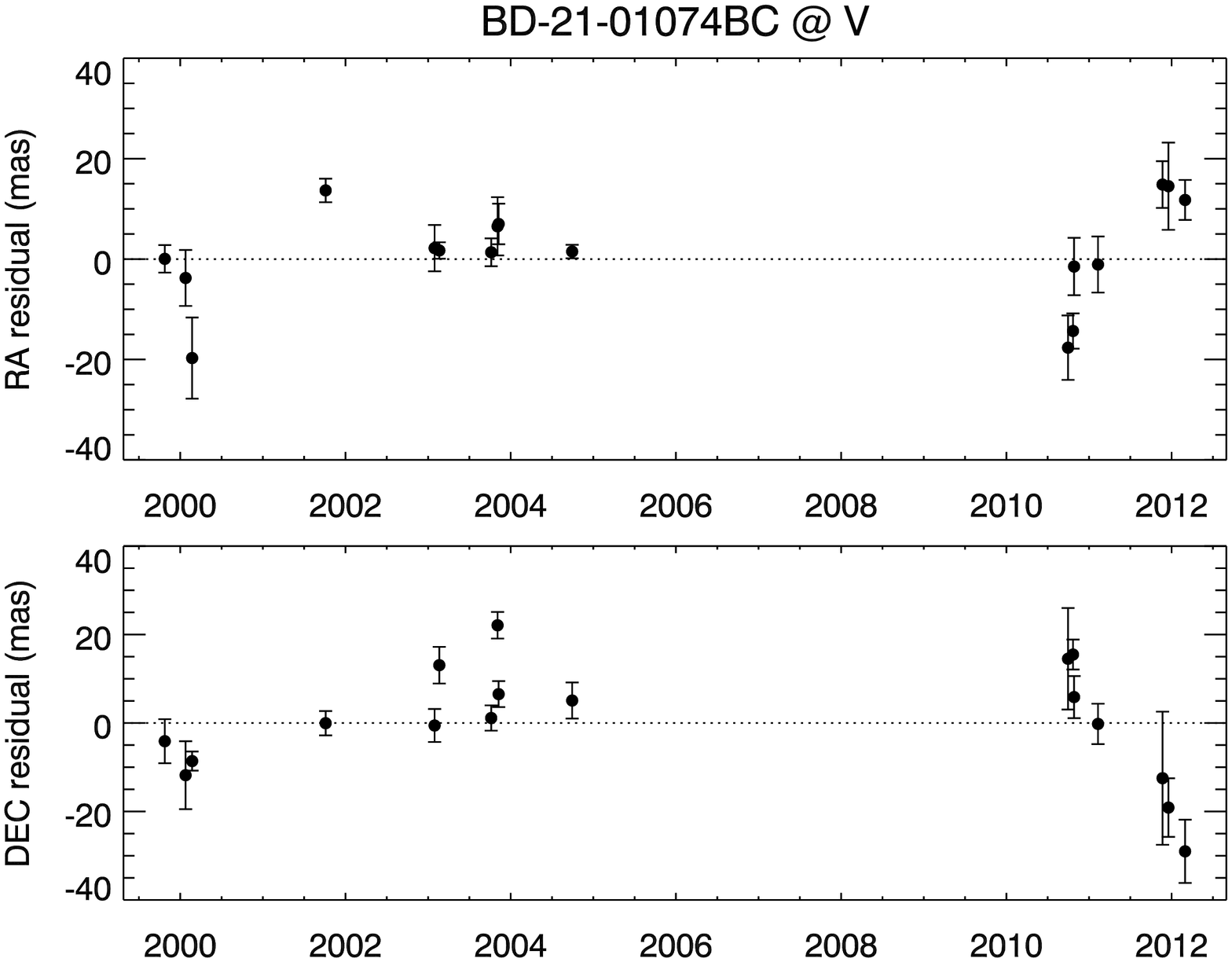}
\caption[BD-21~01074BC astrometric perturbation]{Nightly mean
residuals of the parallax fit (after the parallax and proper motion
have been removed) show a clear astrometric perturbation in both R.A.
and Decl. axes.  This system is a known binary, but the orbit has not
wrapped, and our orbit fit does not converge.
  \label{fig:BD-21-01074BC_orbit}}
\end{figure}

\begin{figure} 
\centering
\includegraphics[angle=90,width=.4\textwidth]{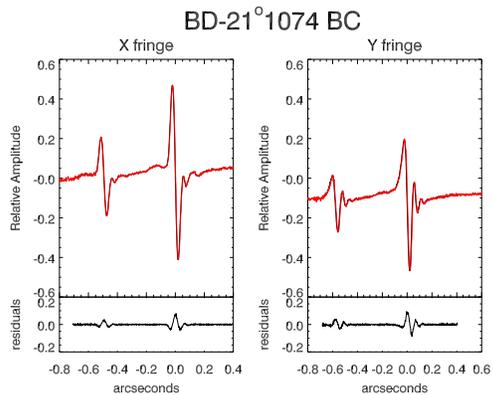}
\caption[BD-21$^{\circ}$1074BC FGS interferogram]{The X-axis (left)
and Y-axis (right) Hubble Space Telescope Fine Guidance Sensor
preliminary results for BD-21$^{\circ}$1074BC. Both axes show the
pronounced presence of a second component (note that the axes shown
here are not R.A. and Decl., but rather the FGS 1r axes at the time of
the observation. The quoted position angles elsewhere have been
corrected for spacecraft roll angle).  Lower panel shows the residuals 
to the fit; the subtraction is not perfect.
  \label{fig:BD-21-01074BC_FGS}}
\end{figure}

\begin{figure} 
\centering
\includegraphics[angle=0,width=.4\textwidth]{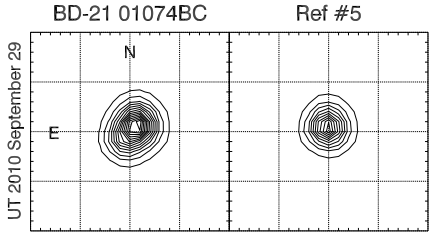}
\caption[BD-21~01074BC]{Contour plot of BD-21$^{\circ}$1074BC on 2010
Sep 29 from CTIOPI data; the SE elongation is probably the C component, despite
WDS claiming the position angle is 321$^{\circ}$.  The nearest reference 
star (\#5) is plotted as a
representative single-star PSF, with 4$\times$ smaller contour
intervals.  BD-21$^{\circ}$1074A is saturated on this frame and not shown.
Grid lines are 2.005\arcsec~apart (5 pixels at the CTIO 0.9m).
  \label{fig:BD-21-01074BC}}
\end{figure}

The BC component was observed by the Hubble Space Telescope Fine Guidance
Sensor Interferometer on 2008 Dec 18, and resolved (Figure 
\ref{fig:BD-21-01074BC_FGS}) in both axes, with details in Table 
\ref{tab:multiples}. The measured position angle is discrepant with 
the current Washington Double Star catalog value
(0.8\arcsec~@ 321$^{\circ}$) but matches the angle of the visible elongation of
the BC point spread function from CTIOPI data, as seen in Figure 
\ref{fig:BD-21-01074BC}.  We see an astrometric binary signal in our 
astrometry, as shown in Figure \ref{fig:BD-21-01074BC_orbit}

\begin{figure} 
\centering
\includegraphics[angle=90,width=.4\textwidth]{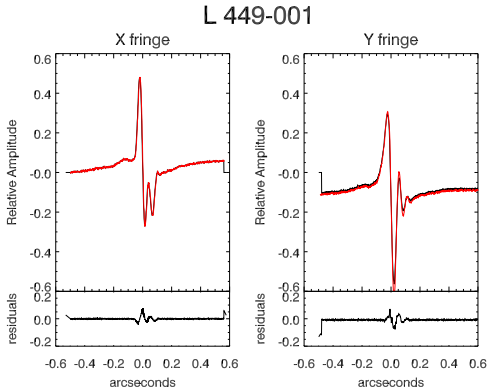}
\caption[L 449-1AB FGS interferogram]{The X-axis (left) and Y-axis
(right) Hubble Space Telescope Fine Guidance Sensor preliminary
results for L~449-1AB.  The X-axis ``S-Curve'' of the Fine Guidance
Sensor shows a second dip to the right of the main one, revealing a
companion.  The residuals to the fit (bottom) demonstrate that the companion 
is not readily resolved in the Y-axis S curve.
  \label{fig:L449-001AB_FGS}}
\end{figure}

{\bf L~449-1AB} was identified as an active star by \citet{Scholz2005}.  
This system was on the HST Cycle 16B FGS proposal, and was resolved into 
two stars on 2008 Dec 03 (Figure \ref{fig:L449-001AB_FGS}) 
with $\Delta F583W$ = 0.95 mag (see Table \ref{tab:multiples}).  By HR 
diagram isochrones and gravity indices, this system is indistinguishable 
from main sequence stars.  The system's kinematics are a potential 
match for the Ursa Major moving group, which given the probable age 
of 500 Myr \citep{King2003} again suggests that the
components may be nearly Zero-Age Main Sequence (ZAMS).

{\bf LHS~1358} has kinematics consistent wih the Hyades stream.  Various 
authorities \citep[e.g.][]{Famaey2008} dispute the existence of the stream 
as a real kinematic entity; at only 15.3$\pm$0.5 pc from the Sun, LHS 1358 is a 
minimum of 30 pc (3 tidal radii) from the Hyades cluster center, and 
therefore the identification is not physical.

\begin{figure} 
\centering
\includegraphics[angle=0,width=.2\textwidth]{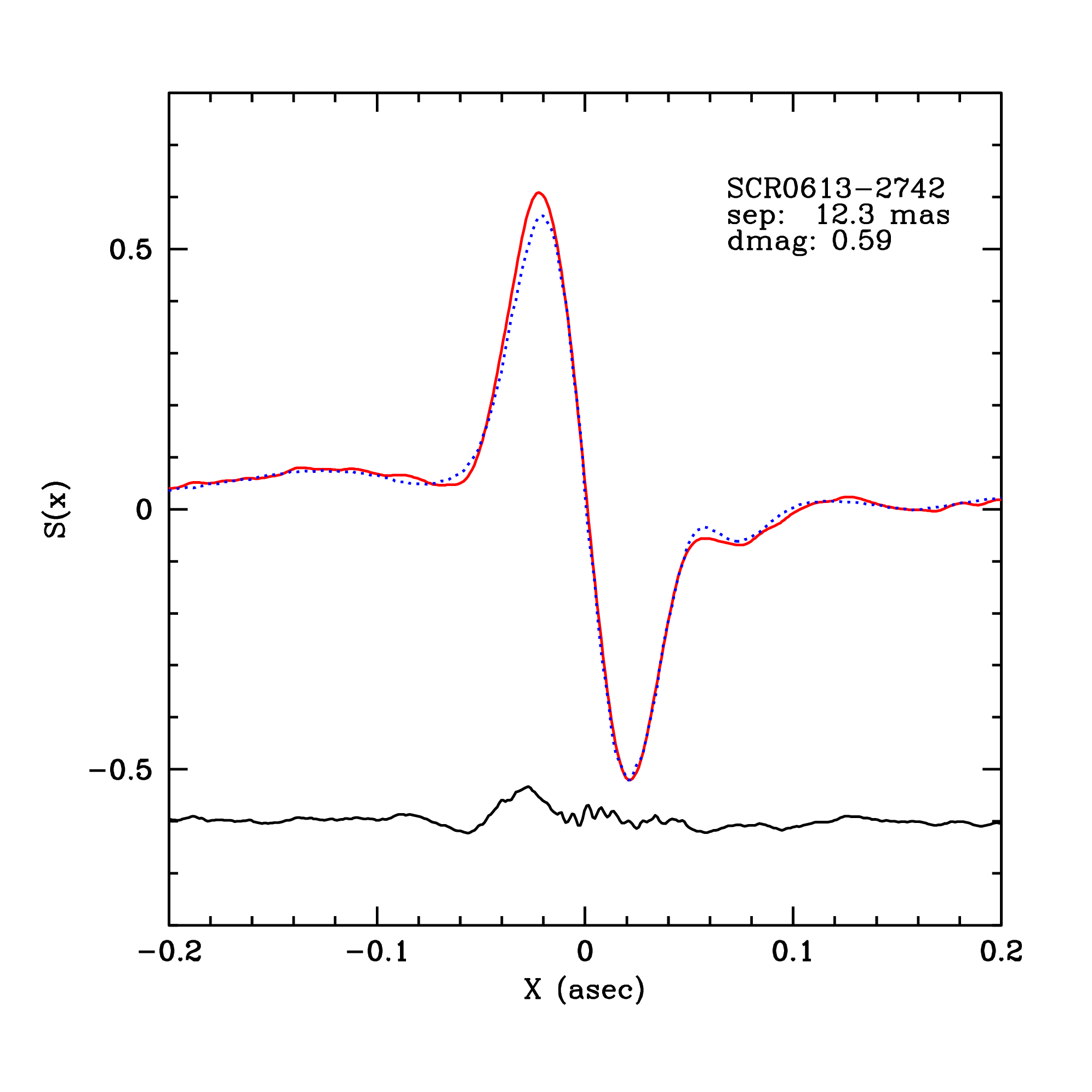}
\includegraphics[angle=0,width=.2\textwidth]{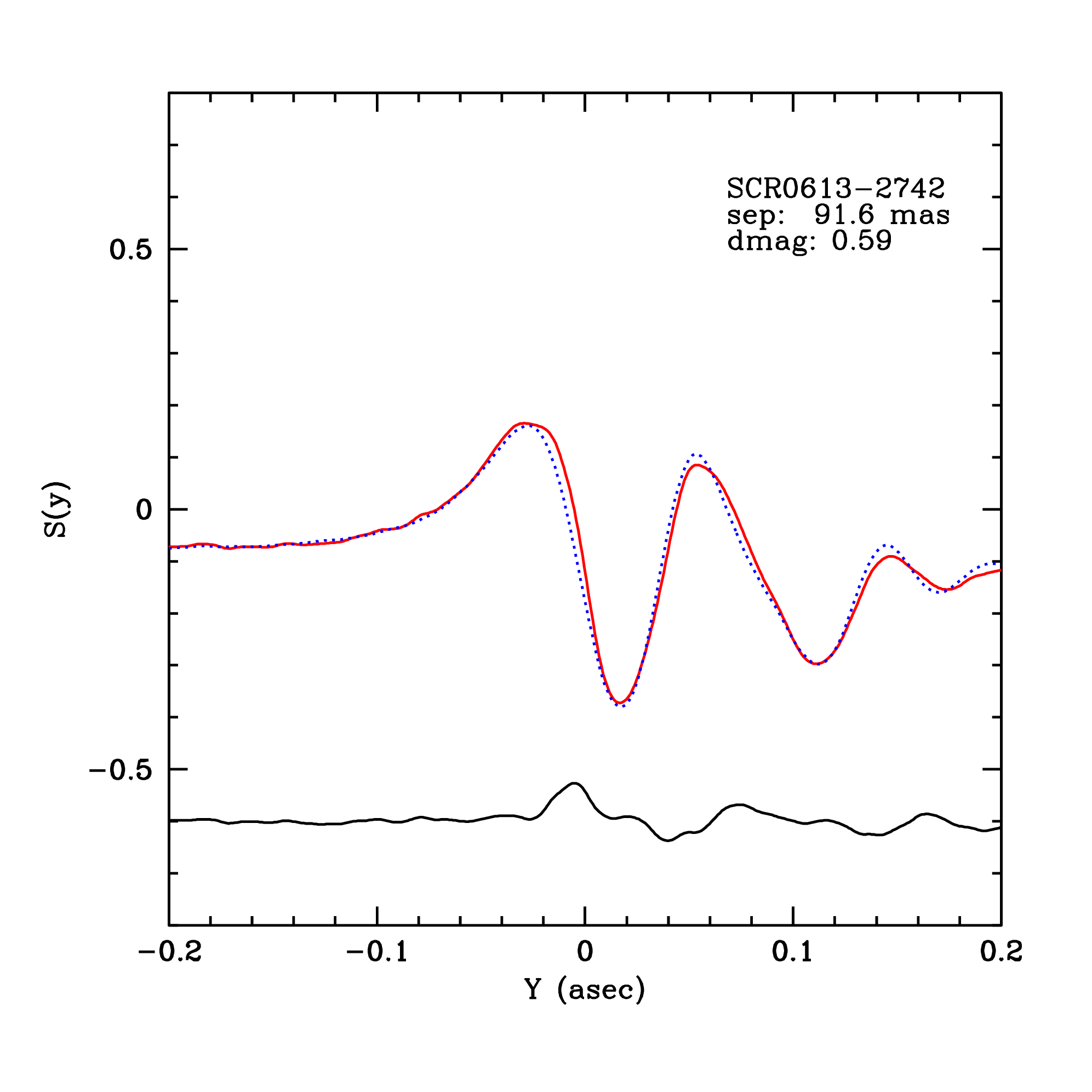}
\caption[SCR~0613-2742AB FGS interferogram]{The X-axis (left) and
Y-axis (right) Hubble Space Telescope Fine Guidance Sensor results for
SCR~0613-2742AB. The Y-axis ``S-curve'' of the Fine Guidance Sensor
shows a second dip to the right of the main one (compare to the axes
of L~449-1AB, Figure \ref{fig:L449-001AB_FGS}), revealing a companion.  The
companion is also barely resolved at $\pm$12 mas (near the limit of
FGS's capabilities) in the X-axis, though this is not visibly
apparent, and carries a sign ambiguity.
  \label{fig:SCR0613-2742AB_FGS}}
\end{figure}

\begin{figure} \centering
\includegraphics[angle=90,width=.4\textwidth]{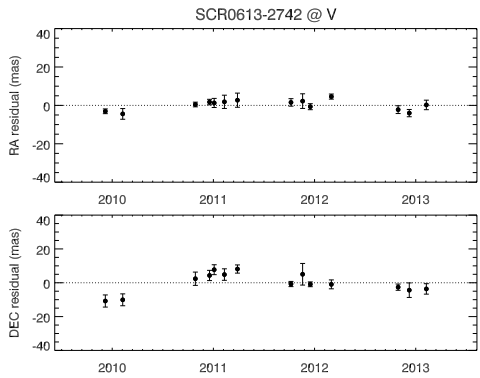}
\caption[SCR 0613-2742AB astrometric perturbation]{Nightly mean
residuals of the parallax fit (after the parallax and proper motion
have been removed) show an astrometric perturbation in both R.A.
and Decl. axes.  We have resolved this binary with FGS, but we do not
yet have an orbit.
  \label{fig:SCR0613-2742_orbit}}
\end{figure}

{\bf SCR~0613-2742AB} is the lowest proper motion system (11.2$\pm$1.0 
mas yr$^{-1}$) thus far observed on CTIOPI, with a transverse velocity 
of 1.6 km s$^{-1}$.  The system is a binary, observed on 2008 Dec 4 and 
resolved into two stars (Figure \ref{fig:SCR0613-2742AB_FGS}, Table 
\ref{tab:multiples}) by FGS, and is also detected as a binary by our FEROS 
spectroscopy.  We see orbital motion in the CTIOPI astrometric residuals 
of this star (Figure \ref{fig:SCR0613-2742_orbit}) and have 
removed that orbital motion\footnote{The orbit that was fit assumes a 4.67 year 
orbit starting at $T_0$ of 2010.803 Julian Years, with photometric semimajor 
axis 0.54 arcsec, inclination 89.9 degrees, eccentricity of 0.9999, a longitude 
of the ascending node of 201.3 degrees, and longitude of periastron of 90.3 
degrees. While it removes the orbital bias from our astrometric data, we do 
not consider the orbit correct.} from our astrometric results 
\citep{Riedel2010}, but we do not have data on a full orbit.  Based on 
the FEROS spectroscopy, we see no Li~6708\AA~doublet absorption. This is 
expected for $\beta$ Pictoris members near the Lithium Depletion 
Boundary \citep{Yee2010}.

The position and proper motion of SCR~0613-2742AB, near the convergent
point of the $\beta$ Pic association, are extremely similar to those of
\object{2MASS J06085283--2753583} \citep{Cruz2003,Rice2010}, a brown
dwarf $\beta$ Pic member.  The parallax of 2M0608 (32.0$\pm$3.6 mas,
 31.3$\pm$3.5 pc) \citep{Faherty2012} is also very similar to SCR0613
 (34.0$\pm$1.0 mas, 29$\pm$0.9 pc).  The two systems are separated
by 3529\arcsec~@78.6$^{\circ}$, which yields a minimum projected
separation of 100 kAU ($\approx$0.5 pc).  It is difficult to compare
their proper motions -- ($+$8.9,$+$10.7)$\pm$(3.5) mas yr$^{-1}$ relative 
(2M0608) and ($-$13.1,$-$0.3)$\pm$(11.6,15.5) mas yr$^{-1}$ absolute 
(SCR0613) -- given the large uncertainties involved, save that each
are marginally consistent with both zero and each other.

Meanwhile, the binding energy, $U_g = {{-GM_1M_2} \over {r}} = {-1.28
\times 10^{33} J}$ using the conservative estimates that the two 
components of SCR0613 have a combined mass of 0.5\msun, and 2M0608 
is 0.015\msun.  This binding energy is lower than any of the extremely 
wide systems studied in \citet{Caballero2009}, and it is probable that these 
stars are not gravitationally bound.  It is possible that this {\it was} 
a bound system at one point (like AU/AT Mic=GJ 803/799AB, below), but 
is not any longer.

{\bf SCR~0757-7714} is overluminous by 2 magnitudes, plotting near the 
$\beta$ Pic isochrone on the HR diagram in Figure \ref{fig:isochrones}, 
and has low surface gravity based on its \ion{K}{1} measurement, but 
not \ion{Na}{1}.  Surprisingly, it shows no other 
signs of youth or membership in any known association -- in fact, it has 
no X-ray emission, and is the only star under consideration with 
H$\alpha$ in {\it absorption}.  Thus, the star's elevation on the HR 
diagram is most likely due to unresolved multiplicity.

{\bf L~34-26} is a potential kinematic match to the Ursa Major moving group, 
but its measured radial velocity ($+$0.9 km s$^{-1}$, no error, \citealt{Torres2006}) 
is most likely discrepant with the best-fit Ursa Major moving group radial 
velocity, $+$6.6 km s$^{-1}$.

\begin{figure} 
\centering
\includegraphics[angle=0,width=.4\textwidth]{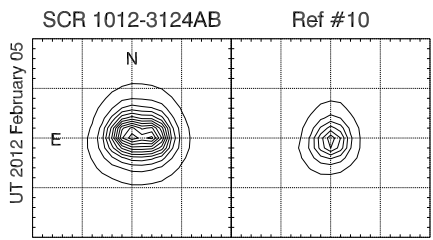}
\caption[SCR 1012-3124AB]{Positions of SCR~1012-3124A (E) and B (W) on
a rare night when the binary is detectable in CTIOPI data.  Reference star \#10 is also
plotted for a comparison of a single-star PSF from the same night,
with the same contour intervals.  Grid lines are 2.005\arcsec~apart 
(5 pixels at the CTIO 0.9m).
  \label{fig:SCR1012-3124AB}}
\end{figure}

{\bf SCR~1012-3124AB} is a close $\sim$1\arcsec~binary (Figure
\ref{fig:SCR1012-3124AB}, Table \ref{tab:multiples}). There are
a few epochs where the B component (to the west of the A component)
can be seen, but for the most part the two components are blended in our 
astrometric observations.

Radial velocity measurements with the VLT-UT1 CRyogenic high-resolution 
infraRed Echelle Spectrograph (CRIRES) presented in Malo et al. {\it submitted}
confirm the star's multiplicity, and yield radial velocities of 14.69$\pm$0.53
km s$^{-1}$ and 14.43$\pm$0.75 km s$^{-1}$, for A and B respectively.  The $v\sin i$
measurements (A: 15.52$\pm$2.01 km s$^{-1}$; B: 20.40$\pm$2.58
km s$^{-1}$) are also indicative of youth, where
\citet{Reiners2009} considers $v\sin i > 20$ km s$^{-1}$ the minimal
condition for a `fast rotator'.

While it is outside the normal spatial bounds \citep{Torres2008} 
of the TW Hydra association (its R.A. is slightly lower than that of TWA~21, 
at 10h13m), its UVW kinematics are consistent with TW
Hydra, as is its deblended isochrone position (Figure 
\ref{fig:isochrones}).  Its gravity measurements (Figure \ref{fig:gravity}) 
indicate it is extremely young.

{\bf TWA 8AB} was listed by \citet{Jao2003} as RXJ1132-264AB and was not
included in the TW Hydra analysis of \citet{Weinberger2013}; our
information is consistent with their conclusions based on other members 
of TW Hydra in that analysis.

{\bf G~165-8AB} appears to be between the ages of $\beta$ Pictoris and 
AB Doradus based on its gravity measurement and deblended HR diagram positions.  
There are two highly discrepant radial velocities published for this 
system, $+8\pm 0.1$ km s$^{-1}$, \citet{Montes2001} and $-7.5\pm 6.5$ km 
s$^{-1}$, \citet{Gizis2002}.  We would normally choose the former value 
due to its higher precision, but the kinematics derived using that value agree with 
no known association.  Ignoring radial velocities altogether, we find 
best-fit matches to to the Tucana-Horologium  ($\gamma$=0.58, best-fit RV $-$6.2 km s$^{-1}$), 
Carina ($\gamma$=0.62, best-fit RV $-$10.4 km s$^{-1}$), and Columba 
($\gamma$=2.40, best-fit RV $-$12.5 km s$^{-1}$) associations, all with 
estimated ages of 30 Myr.  It seems likely that the \citet{Gizis2002} radial 
velocity is accurate, and the uncertainty is accounting for orbital motion 
of the binary. Using the latter radial velocity, the best match 
is to Carina, though BANYAN favors Columba.

Unusually, this system is a northern hemisphere target well outside 
the published spatial boundaries of all three associations.
This suggests that either there is a fourth 30 Myr old association 
nearby, possibly containing two other northern hemisphere objects thought 
to be members of the 30 Myr old Columba association: HR 8799 
\citep{Marois2008,Baines2012} and $\kappa$ And \citep{Carson2013}, or 
that the existing assumptions about the boundaries of the known 
associations are incorrect.

{\bf SCR~1425-4113AB} is a curious system.  High resolution ESPaDOnS 
spectra of this target show that it is a spectroscopic binary, and an 
extremely rapid rotator ($v\sin i =$ 95.0$\pm$8.1 km s$^{-1}$), with a 
radial velocity of ($-$1.2$\pm$1.3 km s$^{-1}$).  Its lithium EW was also 
measured at 595$\pm$20 m\AA, confirming its youth.  The deblended 
magnitudes of both of its components place them above (though consistent with) 
the TW Hydra isochrone on an HR diagram, and the gravity measurements 
(which should not be affected by multiplicity) place it at lower surface 
gravity than TWA 8AB and SCR~1012-3124AB (though still within the errors).

Our kinematics find it to be a much better fit to $\beta$ Pictoris ($\gamma$=0.59) 
than TW Hydra ($\gamma$=6.46), but BANYAN finds 75\% probability of it being 
a TW Hydra member, compared with 24\% probability that it is actually a 
$\beta$ Pictoris member.  It is additionally outside the normal spatial 
boundaries of TW Hydra (being a full hour of R.A. `higher' than the highest 
current member, TWA~18, at 13h21m R.A.).  We nevertheless consider it as a 
potential TW Hydra member because with its extremely high luminosity, low 
surface gravity, and position relatively close to TW Hydra, it either must 
be an outlying member of TW Hydra or something equally young.

\begin{figure} 
\centering
\includegraphics[angle=90,width=.4\textwidth]{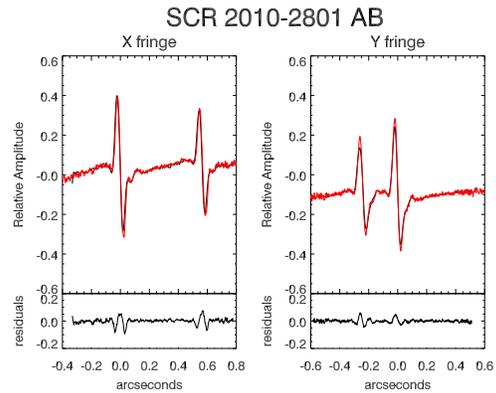}
\caption[SCR~2010-2801AB FGS interferogram]{The X-axis (left) and
Y-axis (right) Hubble Space Telescope Fine Guidance Sensor results for
SCR~2010-2801AB. The dual ``S-curves'' of the two components, originally 
resolved by \citet{Beuzit2004}, are easily visible.  The residuals, on 
the bottom, demonstrate the quality of the fit.
  \label{fig:SCR2010-2801AB_FGS}}
\end{figure}

{\bf SCR~2010-2801AB} was found to be a binary by \citet{Bergfors2010}.  
The system was observed with HST-FGS on 2009 Apr 26, 
as seen in Figure \ref{fig:SCR2010-2801AB_FGS}.  The resulting separation
agrees with the separation published in \citet{Bergfors2010}.

{\bf LEHPM2-0783} has X-ray emission and the strongest 
H-$\alpha$ emission in the sample, despite its cool temperatures.  
It lies above the main sequence, though it is too red to be seen on 
Figure \ref{fig:isochrones}.  It is, however, not a potential kinematic
match to any known young association.  It may be younger than a typical field 
star, but we have no reason to suspect it is anything more than an 
unresolved binary.

{\bf L~755-19} is kinematically consistent with both Castor and
Argus.  It lies within the upper main sequence locus, above the AB
Doradus isochrone, which would be more consistent with the
Argus association (younger than AB Dor) or a main sequence binary than 
the Castor moving group (older than AB Dor).  L~755-19 is a hotter star than 
AP Col \citep{Riedel2011}, and its surface gravity is comparable with either 
field stars or Argus-age objects.  A radial velocity 
would go a long way toward determining the status of this star, given 
that the best-fit Argus RV is $-$25.4 km s$^{-1}$, and the best-fit 
Castor RV is $-$9.3 km s$^{-1}$, and a Castor/Main Sequence assignment 
would imply it is a binary.

{\bf SCR~2033-2556} is consistent with being a member of $\beta$
Pictoris, but its high luminosity suggests it is an unresolved binary.
Observations with ESPaDOnS (Malo et al., {\it in prep}) suggest a
lithium equivalent width of 510 m\AA, confirming the star's youth.

{\bf GJ~803 (A)/GJ~799AB (BC)} is one of the nearest young systems 
(see Table \ref{tab:multiples}) to the Sun, and a prototypical member of
the $\beta$ Pic association \citep{Barrado-y-Navascues1999}.  AU Mic
(unobserved by CTIOPI) is known to have a dust disk, and several
authors note that the AU-AT Mic separation is very large (at least
0.23 pc) and ``must be very fragile and will soon be torn apart by
third bodies'' \citep{Caballero2009}.  This system is a
well-established member of the 10 parsec sample with parallaxes from
{\it HIPPARCOS} and the Yale Parallax Catalog \citep{van-Altena1995}; 
our results agree with the published values.

\begin{figure} 
\centering
\includegraphics[angle=0,width=.4\textwidth]{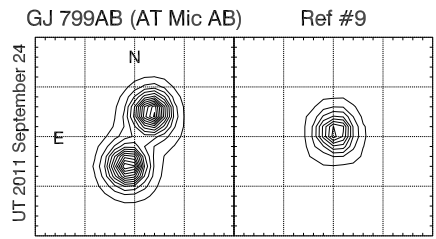}
\caption[GJ 799AB]{GJ 799 A (N) and B (S) on 2011 Sep. 24 from CTIOPI data.  The
nearest reference star (\# 9) is plotted as an example of a
single-star PSF, with 800$\times$ smaller contour intervals.  Grid
spacings are 2.05\arcsec, 5 pixels at the CTIO 0.9m.
  \label{fig:GJ0799AB}}
\end{figure}

In our first epoch (2003 JULY 09) GJ 799 A and B  were separated by 
2.8\arcsec~@ 171$^{\circ}$; in our final epoch (2012 JULY 31, Figure 
\ref{fig:GJ0799AB}) they were separated by 2.3\arcsec~@ 153$^{\circ}$.

{\bf GJ 1284AB} is identified as a potential member of Columba by 
kinematic analysis; however, \citet{Torres2006} identified it as a 
spectroscopic binary, which implies it is a system of main-sequence 
stars, and the kinematic match is spurious.

\section{Conclusions}
\label{sec:conclusions}
We have identified 21 young systems, of which 14 are new, and six 
-- LP~467-16AB, G~7-34, L~449-1AB, G~165-8AB, L~755-19, and SCR~2033-2556 -- are within 25 pc 
of the Sun, as outlined in Table \ref{tab:statisticstable}. This constitutes 6\% of 
all known $\beta$ Pictoris members, and adds two new TW Hydra members that
are outside the currently understood boundaries of the system.  We also
further increase the number of nearby Argus and AB Doradus members.  
Additionally, the enormous number of multiples will be enormously 
beneficial to studies of stellar masses for young stars; there are less than
10 masses for M dwarf stars less than 10 Myr old \citep{Mathieu2007} and 
therefore considerable discrepancies in the various predicted evolutionary 
tracks of low-mass stars; this paper includes three new spectroscopic 
binaries in TW Hydra and $\beta$ Pictoris where {\it every} component 
is less than 0.5\msun.

The most interesting results of this analysis are the contradictions 
with expected young star behavior.  Six systems in the sample are 
clearly overluminous but none of their components have RASS X-ray detections.
Three of them -- SCR~0103-5515ABC, 
SCR~1012-3124AB, and SCR~1425-4113AB -- show H$\alpha$ emission 
in low-resolution optical spectra; all three are definite young systems, 
and all three are multiples.  The fourth target, SCR~0757-7114, shows 
no signs of activity at all and is more likely a triple or quadruple system.  
The remaining two objects -- NLTT~372 and TWA~27AB -- were not observed 
spectroscopically, so we cannot comment on their H$\alpha$ EW (though 
\citet{Gizis2004} recounts wildly varying H$\alpha$ emission from TWA~27AB).

We checked the XMM-Newton and Chandra observing logs to see if any of the six 
systems were observed by higher-sensitivity equipment, and found a Chandra 
observation of TWA~27AB.  That observation formed the basis of the work 
by \citet{Gizis2004}, who report no detected X-ray flux in a 50 ksec 
observation, and provided only an 
upper limit.  Given that TWA~27AB was the only target that might have had 
saturated X-rays given the upper limit from RASS photometry, we can fairly 
conclusively state that none of these systems exhibit X-ray emission.

It is not clear how these six systems (or GJ 2022B and LP 993-115) exhibit 
such low X-ray activity at such young ages, 
except to point out that H$\alpha$ emission is thought to be produced in the 
chromosphere, and X-ray activity originates in the corona; \citet{Riaz2006} 
and \citet{West2009} have noted that H$\alpha$ emission does not correlate 
well with X-ray activity. \citet{Gizis2004} suggest the enormous (and variable) 
H$\alpha$ emission they have compiled from TWA~27AB, combined with its lack 
of X-ray activity, implies that the corona and chromosphere are quiet, and 
the H$\alpha$ activity actually comes from accretion.  Given that 
SCR~1012-3124AB and SCR~1425-4113AB are both suspected members of the 
same ~8 Myr old TW Hydra association, it is not inconceivable for them to 
have disks, though more observations will be needed.  Explaining a disk 
around SCR~0103-5515ABC is more difficult, considering that as a member 
of Tucanae-Horologium it is expected to be $\sim$30 Myr old, though 
with its high luminosity for the two stellar components, we do have evidence 
that it may be a younger system.

The issue of X-ray emission highlights the reason why we favor a 
checklist-style or multi-parameter approach to youth, such as those used 
by \citet{Shkolnik2009,Shkolnik2011}.  None of the indicators of 
youth 
are infallible -- if we were to strictly require 
X-ray saturation, we would have missed both new potential TW Hydra 
members, and have only noticed the AB Doradus member GJ~2022B \citep{Shkolnik2012} 
on the strength of its companions -- but together they can provide a clearer 
and more robust picture of youth.  Equally important, these stars are 
variable, and some number of them will be caught by any given survey in a 
transient inactive state.

We also have a small assortment of stars that do not quite fit into our 
current knowledge of young clusters: Both of the putative new TW Hydra members
(SCR~1012-3124AB and SCR~1425-4113AB) lie outside the spatial bounds of
the TW Hydra association members defined in \citet{Torres2008}; G~165-8AB
is clearly young, but lies outside the spatial and kinematic
boundaries of all known suitably young associations. SCR~0103-5515A
and B are each too luminous to be members of Tucana-Horologium or Carina unless
both are further close multiples. (each lies $\sim$0.4 magnitudes above 
the $\beta$ Pictoris isochrone, as do LP~476-207B, SCR~2010-2801A and B, 
and SCR~0017-6645, though in those cases our deblending method may be 
at fault, see \S\ref{sec:deblending}).  Ultimately, further study is needed
to determine the true extent of these young associations, and determine
what manner of processes have brought the associations to their current state.

\acknowledgements

The RECONS effort is supported primarily by the National Science
Foundation through grants AST 05-07711 and AST 09-08402, and was
supported for a time through NASA's {\it Space Interferometry
Mission}.  Observations were initially made possible by NOAO's Survey
Program and have continued via the SMARTS Consortium.  This research has 
made use of the SIMBAD database and the VizieR catalogue access tool, 
operated at CDS, Strasbourg, France; NASA's Astrophysics Data System;
the SuperCOSMOS Science Archive, prepared and hosted by the Wide Field 
Astronomy Unit, Institute for Astronomy, University of Edinburgh, which 
is funded by the UK Science and Technology Facilities Council; the Washington 
Double Star Catalog maintained at the U.S. Naval Observatory; and the NStars 
project.  This publication makes use of data products from the Two Micron All 
Sky Survey, which is a joint project of the University of Massachusetts and 
the Infrared Processing and Analysis Center/California Institute of Technology, 
funded by the National Aeronautics and Space Administration and the National 
Science Foundation, and the AAVSO Photometric All-Sky Survey (APASS), funded 
by the Robert Martin Ayers Science Fund.  This research has made use of the 
Washington Double Star Catalog maintained at the U.S. Naval Observatory.  
D.R.R. acknowledges support from project BASAL PFB-06 of CONICYT, a Joint 
Committee ESO-Government of Chile grant, and FONDECYT grant \#3130520.  
S.C.B and E.L.R acknowledge support from NASA award 11-ADAP11-0169.

The authors also wish to thank Dr. Brian Mason for supplying the
orbit-fitting code; Dr. Stella Kafka for the CTIO 4.0m spectra, 
Nikhil van der Klaauw, Sean Samaroo, Emmett Goodman-Boyd,
Jonathan Gagn\'e and Dr. Inseok Song for useful discussions, and the 
staff of the Cerro Tololo Inter-American Observatory, particularly 
Edgardo Cosgrove, Arturo Gomez, Alberto Miranda, and Joselino Vasquez, 
for their help over the years.


\bibliographystyle{apj} \bibliography{riedel_a}

\begin{deluxetable}{l|ccc|ccc}
\setlength{\tabcolsep}{0.04in}
\tablewidth{0pt}
\tabletypesize{\scriptsize}
\tablecaption{Young Star Statistics}\label{tab:statisticstable}
\tablehead{
  \colhead{} &
  \multicolumn{3}{c}{All\tablenotemark{a}} &
  \multicolumn{3}{c}{Within 25 pc\tablenotemark{b}}\\

  \colhead{Association}              &
  \colhead{Known members}                  &
  \colhead{M+ dwarfs}                  &
  \colhead{New members}               &
  \colhead{Known members}               &
  \colhead{M+ dwarfs}                  &
  \colhead{New members}           }
\startdata
\hline
$\epsilon$ Chameleontis &  40 & 19 & 0 &  0 &  0 & 0 \\
TW Hydra            &  41 & 31 & 2 &  0 &  0 & 0 \\
$\beta$ Pictoris    & 111 & 68 & 7 & 18 & 16 & 2 \\
Octans              &  15 &  0 & 0 &  0 &  0 & 0 \\
Tucana-Horologium   &  90 & 26 & 0 &  3 &  2 & 0 \\
Columba             &  74 & 20 & 0 &  0 &  0 & 0 \\
Carina              &  34 & 13 & 0 &  4 &  4 & 0 \\ 
Argus               & 123 & 53 & 1 & 25 & 21 & 1 \\
\enddata

\tablecomments{The known members of nearby young associations, with new 
discoveries in this paper.}

\tablenotetext{a}{Members are as defined in the following source papers: \citet{King2003},\citet{Zuckerman2004},\citet{Mamajek2005},\citet{Torres2008},\citet{da-Silva2009},\citet{Shkolnik2009},\citet{Lepine2009},\citet{Rice2010},\citet{Schlieder2010},\citet{Shkolnik2011},\citet{Desidera2011},\citet{Kiss2011},\citet{Riedel2011},\citet{Rodriguez2011},\citet{Zuckerman2011},\citet{McCarthy2012},\citet{Schlieder2012a},\citet{Schlieder2012c},\citet{Shkolnik2012}.  Quality of membership is as decided in the source paper and has not been re-evaluated here.}
\tablenotetext{b}{25 pc is determined by trigonometric parallax if possible; otherwise, from published kinematic, photometric, or spectroscopic distances (in that order)}

\end{deluxetable}

\clearpage

\thispagestyle{empty}

\voffset080pt{
\begin{deluxetable}{llrrrclcccrrrccccl}
\rotate
\setlength{\tabcolsep}{0.02in}
\tablewidth{0pt}
\tabletypesize{\tiny}
\tablecaption{Photometric Results\label{tab:photometry}}
\tablehead{\colhead{}              &
	   \colhead{Alternate}     &
	   \colhead{}              &
 	   \colhead{}              &
	   \colhead{}              &
	   \colhead{No. of phot.}  &
	   \colhead{}              &
	   \colhead{$\sigma$}      &
	   \colhead{No. of rel.}   &
	   \colhead{No. of}        &
	   \colhead{$J$}           &
	   \colhead{$H$}           &
	   \colhead{$K_s$}         &
	   \colhead{spectral}      &
	   \colhead{}              &
	   \colhead{phot.}          &
	   \colhead{No. of}        &
	   \colhead{}              \\

           \colhead{Name}          &
           \colhead{Name}          &
	   \colhead{$V_{J}$}       &
 	   \colhead{$R_{KC}$}      &
	   \colhead{$I_{KC}$}      &
	   \colhead{Observations} &
	   \colhead{$\pi$ filter}  &
	   \colhead{(mag)}         &
 	   \colhead{Nights}        &
 	   \colhead{Frames}        &
	   \colhead{(2MASS)}       &
	   \colhead{(2MASS)}       &
	   \colhead{(2MASS)}       &
	   \colhead{type\tablenotemark{b}}&
	   \colhead{ref}           &
	   \colhead{dist. (pc)}    &
	   \colhead{Relations}     &
	   \colhead{Notes}         \\

           \colhead{(1)}           &         
           \colhead{(2)}           &
           \colhead{(3)}           &
           \colhead{(4)}           &
           \colhead{(5)}           &
           \colhead{(6)}           &
           \colhead{(7)}           &
           \colhead{(8)}           &
           \colhead{(9)}           &
           \colhead{(10)}          &
           \colhead{(11)}          &
           \colhead{(12)}          &
           \colhead{(13)}          &
           \colhead{(14)}          &
           \colhead{(15)}          &
           \colhead{(16)}          &
           \colhead{(17)}          &
           \colhead{(18)}          }

\startdata
\hline
NLTT 372       & LP~404-32 & 15.87$\pm$0.04 & 14.66$\pm$0.03 & 13.12$\pm$0.05 & 3 & $V$ & 0.014 & 21 &  97 & 11.64$\pm$0.02 & 11.00$\pm$0.03 & 10.72$\pm$0.02 & M4.5V   & 1 & 47.39$\pm$7.47 & 12 & \tablenotemark{a}\\
G 131-26AB     & NLTT~375  & 13.52$\pm$0.06 & 12.19$\pm$0.04 & 10.50$\pm$0.04 & 3 & $V$ & 0.019 & 21 &  97 &  8.87$\pm$0.03 &  8.26$\pm$0.03 &  8.01$\pm$0.02 & M4.0VeJ & 2 & 10.74$\pm$1.67 & 12 & \tablenotemark{a}\\
SCR 0017-6645  & RBS~38    & 12.45$\pm$0.02 & 11.37$\pm$0.01 & 10.00$\pm$0.01 & 2 & $V$ & 0.035 & 11 &  54 &  8.56$\pm$0.02 &  7.93$\pm$0.04 &  7.70$\pm$0.02 & M2.5Ve  & 2 & 14.82$\pm$2.69 & 12 & \\
GJ 2006A       & RBS~67    & 12.95$\pm$0.02 & 11.79$\pm$0.03 & 10.29$\pm$0.03 & 3 & $V$ & 0.077 & 17 &  69 &  8.88$\pm$0.03 &  8.24$\pm$0.04 &  8.01$\pm$0.03 & M3.5Ve  & 2 & 15.39$\pm$2.41 & 12 & \tablenotemark{a}\\
GJ 2006B       &           & 13.25$\pm$0.03 & 12.04$\pm$0.02 & 10.48$\pm$0.02 & 4 & $V$ & 0.036 & 17 &  69 &  8.97$\pm$0.03 &  8.39$\pm$0.03 &  8.12$\pm$0.03 & M3.5Ve  & 2 & 14.29$\pm$2.19 & 12 & \tablenotemark{a}\\
SCR 0103-5515ABC&          & 15.48$\pm$0.03 & 14.00$\pm$0.02 & 12.07$\pm$0.02 & 2 & $R$ & 0.021 & 11 &  49 & 10.16$\pm$0.02 &  9.58$\pm$0.03 &  9.24$\pm$0.02 & M4.5Ve  & 2 & 12.86$\pm$1.97 & 12 & \\
LP 467-16ABC   &           & 14.46$\pm$0.05 & 12.95$\pm$0.01 & 11.03$\pm$0.02 & 2 & $R$ & 0.019 & 15 &  79 &  9.08$\pm$0.03 &  8.51$\pm$0.04 &  8.21$\pm$0.03 & M4.5VeJ & 2 &  7.88$\pm$1.22 & 12 & \\
GJ 2022AC      &G~269-153AB& 13.56$\pm$0.04 & 12.26$\pm$0.01 & 10.62$\pm$0.05 & 3 & $R$ & 0.067 & 14 &  31 &  9.20$\pm$0.04 &  8.66$\pm$0.05 &  8.24$\pm$0.03 & M4.0VeJ & 2 & 14.78$\pm$3.01 & 12 & \\
\enddata

\tablecomments{Photometry data collected on the sample.}

\tablenotetext{a}{Astrometric results and relative photometry use new
$V$ filter data.}

\tablenotetext{b}{References: (1) \citet{Reid1995}; (2) This paper; (3) \citet{Gizis2007}.  ``J'' indicates joint spectral types from unresolved multiples.}

\tablenotetext{c}{No independent photometry.}

\end{deluxetable}
}

\clearpage

\voffset80pt{
\begin{deluxetable}{lcccccccccrrcccl}
\rotate
\setlength{\tabcolsep}{0.02in}
\tablewidth{0pt}
\tabletypesize{\scriptsize}
\tablecaption{Astrometric Results\label{tab:astrometry}}
\tablehead{\colhead{}              &
	   \colhead{R.A.}          &
 	   \colhead{Decl.}         &
	   \colhead{}              &
	   \colhead{}              &
	   \colhead{}              &
	   \colhead{}              &
	   \colhead{}              &
	   \colhead{}              &
	   \colhead{$\pi$(rel)}    &
	   \colhead{$\pi$(corr)}   &
	   \colhead{$\pi$(abs)}    &
	   \colhead{$\mu$}         &
	   \colhead{P.A.}          &
	   \colhead{V$_{tan}$}      &
	   \colhead{}              \\

           \colhead{Name}          &
	   \multicolumn{2}{c}{(J2000)\tablenotemark{a}} &
	   \colhead{Filter}        &
	   \colhead{N$_{sea}$\tablenotemark{b}}&
	   \colhead{N$_{frm}$}     &
	   \colhead{Coverage\tablenotemark{b}}&
	   \colhead{Years\tablenotemark{b}}&
 	   \colhead{N$_{ref}$}     &
	   \colhead{(mas)}         &
	   \colhead{(mas)}         &
	   \colhead{(mas)}         &
	   \colhead{(mas yr$^{-1}$)}&
	   \colhead{(deg)}         &
	   \colhead{(km s$^{-1}$)} &
	   \colhead{Notes}         \\
           \colhead{(1)}           &
           \colhead{(2)}           &
           \colhead{(3)}           &
           \colhead{(4)}           &
           \colhead{(5)}           &
           \colhead{(6)}           &
           \colhead{(7)}           &
           \colhead{(8)}           &
           \colhead{(9)}           &
           \colhead{(10)}          &
           \colhead{(11)}          &
           \colhead{(12)}          &
           \colhead{(13)}          &
           \colhead{(14)}          &
           \colhead{(15)}          &
           \colhead{(16)}          }

\startdata
\hline
\object{NLTT 372}        & 00 08 51.79 & $+$20 49 07.9 & $V$ &11s &  97 & 1999.71--2011.89 & 12.18 &  7 &  10.50$\pm$1.44 & 1.16$\pm$0.07 &  11.66$\pm$1.44 & 147.8$\pm$0.3 & 121.1$\pm$ 0.23 & 60.1 & \tablenotemark{c}\\
\object{G 131-26AB}      & 00 08 53.92 & $+$20 50 25.4 & $V$ &11s &  97 & 1999.71--2011.89 & 12.18 &  7 &  52.97$\pm$1.35 & 1.16$\pm$0.07 &  54.13$\pm$1.35 & 251.2$\pm$0.3 & 194.4$\pm$ 0.11 & 22.0 & \tablenotemark{c}\\
\object{SCR 0017-6645}   & 00 17 23.53 & $-$66 45 12.5 & $V$ & 4s &  54 & 2009.75--2012.88 &  3.13 &  9 &  24.71$\pm$1.73 & 0.90$\pm$0.10 &  25.61$\pm$1.73 &  93.6$\pm$1.8 & 101.1$\pm$ 1.82 & 17.3 & \\
\object{GJ 2006A}        & 00 27 50.24 & $-$32 33 06.1 & $V$ & 9s &  69 & 2000.57--2010.82 & 10.25 &  7 &  29.49$\pm$2.50 & 0.65$\pm$0.11 &  30.14$\pm$2.50 & 115.0$\pm$0.6 & 109.6$\pm$ 0.55 & 18.1 & \tablenotemark{c}\\
\object{GJ 2006B}        & 00 27 50.36 & $-$32 33 23.9 & $V$ & 9s &  69 & 2000.57--2010.82 & 10.25 &  7 &  31.13$\pm$2.47 & 0.65$\pm$0.11 &  31.78$\pm$2.47 & 115.8$\pm$0.6 & 108.6$\pm$ 0.53 & 17.3 & \tablenotemark{c}\\
\object{SCR 0103-5515ABC} & 01 03 35.63 & $-$55 15 56.2 & $R$ & 6s &  49 & 2007.82--2012.87 &  5.05 &  8 &  20.77$\pm$1.37 & 0.41$\pm$0.07 &  21.18$\pm$1.37 &  89.0$\pm$1.0 & 112.0$\pm$ 1.24 & 19.9 & \\
\object{LP 467-16AB}     & 01 11 25.41 & $+$15 26 21.6 & $R$ & 8s &  79 & 1999.71--2009.57 &  9.86 &  7 &  44.88$\pm$1.77 & 0.91$\pm$0.16 &  45.79$\pm$1.78 & 222.9$\pm$0.6 & 122.7$\pm$ 0.29 & 23.1 & \\
\object{GJ 2022AC}       & 01 24 30.62 & $-$33 55 01.6 &\ldots&\ldots&\ldots&    \ldots    &\ldots &\ldots&     \ldots    &    \ldots     &       \ldots    &     \ldots    &      \ldots     &\ldots& \tablenotemark{d}\\
\enddata


\tablenotetext{a}{coordinates are epoch and equinox 2000; each
target's coordinates were extracted from 2MASS and then transformed to
epoch 2000 using the proper motions and position angles listed
here.}

\tablenotetext{b}{`Coverage' and `Years' run from the first to last
data point; `Seasons' counts observing semesters where a dataset was
taken, and denotes if coverage was `c'ontinuous (more than one night
of data in all seasons) or `s'cattered.}

\tablenotetext{c}{Astrometric results include new $V$ filter data.}

\tablenotetext{d}{Not using independent astrometry (see $\S$\ref{sec:results})}

\tablenotetext{e}{The astrometric perturbation was incorporated into the
final astrometric fit.}

\tablenotetext{f}{Reference field was reddened, and a generic correction to absolute parallax was adopted.}

\end{deluxetable}
}

\voffset000pt{
\begin{deluxetable}{l|llrc|rrccc}
\setlength{\tabcolsep}{0.02in}
\tablewidth{0pt}
\tabletypesize{\scriptsize}
\tablecaption{Data used to calculate UVWXYZ\label{tab:kinematicstable}}
\tablehead{
  \colhead{Name}        &
  \colhead{R.A.}        &
  \colhead{Decl.}       &
  \colhead{$\pi$}       &
  \colhead{$\pi$}       &
  \colhead{$\mu_{R.A.\cos{Decl.}}$}  &
  \colhead{$\mu_{Decl.}$} &
  \colhead{$\mu$}        &
  \colhead{R.V.}        &
  \colhead{R.V.}         \\

  \colhead{}                &
  \multicolumn{2}{c}{(J2000$\pm$mas)} &
  \colhead{(mas)}           &
  \colhead{Ref.}            &
  \multicolumn{2}{c}{(mas yr$^{-1}$)} &
  \colhead{Ref.}            &
  \colhead{(km s$^{-1}$)}    &
  \colhead{Ref.}            \\

  \colhead{(1)}             &
  \colhead{(2)}             &
  \colhead{(3)}             &
  \colhead{(4)}             &
  \colhead{(5)}             &
  \colhead{(6)}             &
  \colhead{(7)}             &
  \colhead{(8)}             &
  \colhead{(9)}             &
  \colhead{(10)}            }

\startdata
\hline 
NLTT 372           & 00 08 51.788$\pm$ 70 & $+$20 49 07.86$\pm$ 80 &  11.66$\pm$1.44 & 1       &  $+$120.7$\pm$12.1 &  $-$76.0$\pm$9.4 & 1 & \ldots              & \ldots\\
G 131-26AB         & 00 08 53.922$\pm$ 70 & $+$20 50 25.45$\pm$ 80 &  54.13$\pm$1.35 & 1       &   $-$68.5$\pm$12.1 & $-$243.0$\pm$9.4 & 2 & \ldots              & \ldots\\
SCR 0017-6645      & 00 17 23.524$\pm$ 80 & $-$66 45 12.46$\pm$110 &  25.61$\pm$1.73 & 1       &  $+$102.9$\pm$ 1.0 &  $-$15.0$\pm$1.0 & 2 &  $+$11.4 $\pm$ 0.8  & 12 \\
GJ 2006A           & 00 27 50.242$\pm$ 70 & $-$32 33 06.17$\pm$ 70 &  30.97$\pm$1.76 & 1 1     &   $+$99.2$\pm$ 1.3 &  $-$61.3$\pm$2.6 & 2 &   $+$8.4            & 13 \\
GJ 2006B           & 00 27 50.362$\pm$ 70 & $-$32 33 23.91$\pm$ 70 &  30.97$\pm$1.76 & 1 1     &  $+$117.2$\pm$ 4.1 &  $-$31.5$\pm$5.8 & 2 & \ldots              & \ldots\\
SCR 0103-5515ABC   & 01 03 35.635$\pm$ 60 & $-$55 15 56.19$\pm$ 80 &  21.18$\pm$1.37 & 1       &  $+$100.2$\pm$ 2.0 &  $-$47.0$\pm$2.4 & 2 & \ldots              & \ldots\\
LP 467-16AB        & 01 11 25.412$\pm$ 70 & $+$15 26 21.62$\pm$ 60 &  45.79$\pm$1.78 & 1       &  $+$181.8$\pm$12.1 & $-$120.0$\pm$9.4 & 1 &   $+$4.0 $\pm$ 0.1  & 14 \\
GJ 2022A           & 01 24 27.693$\pm$ 60 & $-$33 55 08.75$\pm$ 60 &  39.50$\pm$1.28 & 1 3     &  $+$163.3$\pm$12.1 & $-$125.0$\pm$9.5 & 1 &  $+$18.4 $\pm$ 1.0  &  3 \\
\enddata

\tablecomments{References in this table: 1: This paper, 2: UCAC4 \citep{Zacharias2013}, 3: \citet{Shkolnik2012}, 4: YPC \citep{van-Altena1995}, 
5: HIPPARCOS \citep{van-Leeuwen2007}, 6: \citet{Henry2006}, 7: \citet{Riedel2010}, 8: \citet{Riedel2011}, 9: \citet{Gizis2007}
10: \citet{Biller2007}, 11: \citet{Ducourant2008}, 12: \citet{Malo2013}, 13: Malo et al. {\it in preparation},  14: \citet{Montes2001}, 
15: \citet{Gizis2002}, 16: \citet{Torres2006}, 17: \citet{Shkolnik2011}, 18: \citet{Rice2010}, 19: \citet{Shkolnik2010}. 
Parallaxes of multiple components in the same system have been combined, and are represented by duplicate references in the 
parallax column.  All positions and position errors are taken from 2MASS \citep{Skrutskie2006}, and adjusted to epoch 2000.0 
equinox J2000 using the proper motions listed here. }

\end{deluxetable}
}


\thispagestyle{empty}

\voffset-040pt{
\begin{deluxetable}{lrrrr|rrc}
\setlength{\tabcolsep}{0.02in}
\tablewidth{0pt}
\tabletypesize{\scriptsize}
\tablecaption{Deblended magnitudes for isochrones\label{tab:isochronetable}}
\tablehead{
  \colhead{Name}            &
  \colhead{$V$}             &
  \colhead{$I$}             &
  \colhead{$J$}             &
  \colhead{$K$}             &
  \colhead{$M_V$}           &
  \colhead{$V-K$}           &
  \colhead{Deblend\tablenotemark{a}}         \\

  \colhead{(1)}             &
  \colhead{(2)}             &
  \colhead{(3)}             &
  \colhead{(4)}             &
  \colhead{(5)}             &
  \colhead{(6)}             &
  \colhead{(7)}             &
  \colhead{(8)}             }
\startdata
\hline 
NLTT 372         & 15.87 & 13.12 & 11.64 & 10.72 & 11.20 &  5.15 & \\
G 131-26A        & 13.94 & 10.98 &  9.40 &  8.81 & 12.61 &  5.13 & $K$\\
G 131-26B        & 14.76 & 11.61 &  9.90 &  9.27 & 13.43 &  5.49 & $K$\\
SCR 0017-6645    & 12.45 & 10.00 &  8.56 &  7.70 &  9.49 &  4.75 & \\
GJ 2006A         & 12.95 & 10.29 &  8.88 &  8.01 & 10.40 &  4.94 & \\
GJ 2006B         & 13.25 & 10.48 &  8.97 &  8.12 & 10.70 &  5.13 & \\
SCR 0103-5515A   & 16.07 & 12.70 & 10.83 &  9.91 & 12.70 &  6.16 & $K$\\
SCR 0103-5515B   & 16.43 & 12.97 & 11.04 & 10.11 & 13.06 &  6.32 & $K$\\
\enddata

\tablenotetext{a}{Band of known $\Delta$ mag used to deblend the photometry, or SB if system was identified only as a spectroscopic binary.  No entry is given for single stars because deblending is not required.}

\end{deluxetable}
}

\begin{deluxetable}{l|rrccccrrr}
\setlength{\tabcolsep}{0.04in}
\tablewidth{0pt}
\tabletypesize{\scriptsize}
\tablecaption{Photometric and Spectroscopic Properties\label{tab:youngproperties}}
\tablehead{
  \colhead{Name}                  &
  \colhead{$M_V$}                 &
  \colhead{$V-K_S$}               &
  \colhead{$\log{\frac{L_{X}}{L_{bol}}}$}&
  \colhead{$\log{L_{X}}$}         &
  \multicolumn{2}{c}{Variability} &
  \colhead{H-$\alpha$}            &
  \colhead{\ion{Na}{1}}            &
  \colhead{\ion{K}{1}}             \\
 
  \colhead{}                      &
  \colhead{(mag)}                 &
  \colhead{(mag)}                 &
  \colhead{}                      &
  \colhead{(erg $^{-1}$)}         &
  \colhead{filter}                &
  \colhead{(mag)}                 &
  \colhead{EW \AA}                &
  \colhead{Index}                 &
  \colhead{EW \AA}                \\

  \colhead{(1)}                   &
  \colhead{(2)}                   &
  \colhead{(3)}                   &
  \colhead{(4)}                   &
  \colhead{(5)}                   &
  \colhead{(6)}                   &
  \colhead{(7)}                   &
  \colhead{(8)}                   &
  \colhead{(9)}                   &
  \colhead{(10)}                   }

\startdata
\hline
NLTT 372          & 11.20 & 5.15 &($-$2.22)&(29.73)& $V$ & 0.014 &       &      &     \\
G 131-26AB        & 12.19 & 5.51 & $-$3.28 & 28.40 & $V$ & 0.019 &  -3.5 & 1.25 & 2.2 \\
SCR 0017-6645     &  9.49 & 4.75 & $-$3.01 & 29.49 & $V$ & 0.035 &  -4.8 & 1.11 & 0.5 \\
GJ 2006A          & 10.40 & 4.94 &\multirow{2}{*}{$-$2.96} &\multirow{2}{*}{29.50}& $V$ & 0.077 &  -5.6 & 1.13 & 0.6 \\
GJ 2006B          & 10.70 & 5.13 & & & $V$ & 0.036 & -10.0 & 1.15 & 1.1 \\
SCR 0103-5515ABC  & 12.11 & 6.24 &($-$4.57)&(27.40)& $R$ & 0.021 & -14.5 & 1.25 & 3.0 \\
LP 467-16AB       & 12.76 & 6.25 & $-$3.08 & 28.62 & $R$ & 0.019 &  -9.9 & 1.19 & 2.5 \\
GJ 2022AC         & 11.54 & 5.32 & $-$2.70 & 29.08 & $R$ & 0.067 &  -9.4 & 1.23 & 2.1 \\
\enddata

\tablecomments{Errors on X-ray values are less than 25\%.  Errors on
  spectroscopic EW measures are roughly 0.2\AA.  Errors on the \ion{Na}{1} index are 
  0.05.  Values in parentheses are upper limits.  Three systems have multiple resolved 
  components within the RASS positional uncertainties, and have combined X-ray luminosities}

\end{deluxetable}

\voffset080pt{
\begin{deluxetable}{l|llr|lc|l|l|l|ll}
\setlength{\tabcolsep}{0.04in}
\tablewidth{0pt}
\rotate
\tabletypesize{\scriptsize}
\tablecaption{Young Star Results\label{tab:youngresults}}
\tablehead{\colhead{Name}                  &
           \multicolumn{3}{c}{Kinematic}   &
           \multicolumn{2}{c}{BANYAN}      &
           \colhead{Isochrone}             &
           \colhead{Gravity}               &
           \colhead{H$\alpha$ age}         &
           \colhead{Association}           &
           \colhead{Other}                 \\

         \colhead{}                        &
         \colhead{Match}                   &
         \colhead{$\gamma$}                &
         \colhead{R.V.}                    &
         \colhead{Match}                   &
         \colhead{Prob (\%)}               &
         \colhead{Range}                   &
         \colhead{Match}                   &
         \colhead{(Gyr)}                   &
         \colhead{}                        &
         \colhead{References}              \\

         \colhead{(1)}                     &
         \colhead{(2)}                     &
         \colhead{(3)}                     &
         \colhead{(4)}                     &
         \colhead{(5)}                     &
         \colhead{(6)}                     &
         \colhead{(7)}                     &
         \colhead{(8)}                     &
         \colhead{(9)}                     &
         \colhead{(10)}                    &
         \colhead{(11)}                    }

\startdata
\hline
NLTT 372          &             &      &         & Field       & 99.9  & ABD--$\beta$Pic & N/A   & N/A    & & \\
G 131-26AB        &             &      &         & Field       & 99.9  &Field--ABD & Old   & $<$4.5 & & \\
SCR 0017-6645     & $\beta$ Pic & 0.24 & $+$11.0 & $\beta$ Pic & 99.9  &$\beta$Pic & Young & $<$1.6 & $\beta$ Pictoris & \citet{Riedel2012}, \citet{Malo2013} \\
GJ 2006AB         & $\beta$ Pic & 1.86 & $+$11.4\tablenotemark{a} & $\beta$ Pic & 99.9  &$\beta$Pic & Young & $<$3.0 & $\beta$ Pictoris & \citet{Riedel2012} \\
SCR 0103-5515ABC  & Carina      & 1.68 & $+$14.3\tablenotemark{a} & Tuc-Hor     & 98.9  & TWA--$\beta$Pic & Young & $<$6.0 & Unknown & (\citealt{Delorme2012} Tucana-Horologium)\\
LP 467-16AB       & $\beta$ Pic & 0.46 &  $+$4.0 & $\beta$ Pic & 99.0  &$\beta$Pic--Field& Young & $<$6.0 & $\beta$ Pictoris & \citet{Riedel2012}, \citet{Malo2013} \\
GJ 2022ABC        & AB Dor      & 0.09 & $+$18.4 & AB Dor      & 99.9  & Field     & Young & $<$4.5 & AB Doradus & \citet{Riedel2012}, \citet{Shkolnik2012} \\
LP 993-115ABC     &             &      &         & Field       & 99.9  & Field     & Old   & $<$3.0 & & \\
\enddata

\tablenotetext{a}{R.V. is actually a best-fit R.V. from the kinematic fitting.}

\end{deluxetable}
}

\voffset040pt{
\begin{deluxetable}{lrrrrrrrrrrrrr} 
\setlength{\tabcolsep}{0.04in}
\tablewidth{0pt} 
\tabletypesize{\scriptsize} 
\tablecaption{Nearby Young Associations\label{tab:associationstable}}
\tablehead{
 \colhead{Name} &
 \colhead{U} &
 \colhead{$\sigma_U$} &
 \colhead{V} &
 \colhead{$\sigma_V$} &
 \colhead{W} &
 \colhead{$\sigma_W$} &
 \colhead{min X} &
 \colhead{max X} &
 \colhead{min Y} &
 \colhead{max Y} &
 \colhead{min Z} &
 \colhead{max Z} &
 \colhead{Age} \\

 \colhead{} &
 \colhead{(km s$^{-1}$)} &
 \colhead{(km s$^{-1}$)} &
 \colhead{(km s$^{-1}$)} &
 \colhead{(km s$^{-1}$)} &
 \colhead{(km s$^{-1}$)} &
 \colhead{(km s$^{-1}$)} &
 \colhead{(pc)} &
 \colhead{(pc)} &
 \colhead{(pc)} &
 \colhead{(pc)} &
 \colhead{(pc)} &
 \colhead{(pc)} &
 \colhead{(Myr)} \\

 \colhead{(1)} &
 \colhead{(2)} &
 \colhead{(3)} &
 \colhead{(4)} &
 \colhead{(5)} &
 \colhead{(6)} &
 \colhead{(7)} &
 \colhead{(8)} &
 \colhead{(9)} &
 \colhead{(10)} &
 \colhead{(11)} &
 \colhead{(12)} &
 \colhead{(13)} &
 \colhead{(14)} }

\startdata
 $\epsilon$ Cha &-11.0 & 1.2 & -19.9 & 1.2 & -10.4 & 1.6 &  +34 & +60 & -105 & -78 & -44 & -12 & 6\tablenotemark{a} \\
 TW Hya         &-10.5 & 0.9 & -18.0 & 1.5 &  -4.9 & 0.9 &   +2 & +34 &  -61 & -26 & +10 & +27 & 8\tablenotemark{a} \\
 $\beta$ Pic    &-10.1 & 2.1 & -15.9 & 0.8 &  -9.2 & 1.0 &  -32 & +76 &  -33 &  21 & -29 &  -1 & 12\tablenotemark{a} \\
 Octans         &-14.5 & 0.9 &  -3.6 & 1.6 & -11.2 & 1.4 &  -79 &+142 & -138 & -60 & -85 & -38 & 20 \\
 Tuc-Hor        & -9.9 & 1.5 & -20.9 & 0.8 &  -1.4 & 0.9 &  -61 & +43 &  -47 &  -4 & -44 & -30 & 30 \\
 Columba        &-13.2 & 1.3 & -21.8 & 0.8 &  -5.9 & 1.2 & -106 &  +9 & -168 &  +1 & -99 &  +6 & 30 \\
 Carina         &-10.2 & 0.4 & -23.0 & 0.8 &  -4.4 & 1.5 &   -2 & +33 & -154 & -39 & -33 &  +5 & 30 \\
 Argus          &-22.0 & 0.3 & -14.4 & 1.3 &  -5.0 & 1.3 &  -55 & +64 & -154 &  -6 & -67 &  +8 & 50 \\
\enddata

\tablecomments{All data (unless otherwise specified) from \citet{Torres2008}.}

\tablenotetext{a}{Only these ages are known with any degree of certainty or corroboration.}

\tablenotetext{b}{Age from \citet{Luhman2005}}

\tablenotetext{c}{Dimensions \citet{Soderblom2005}, 13 pc tidal radius
\citet{Adams2001}}

\tablenotetext{d}{\citet{Barrado-y-Navascues1998}}

\tablenotetext{e}{\citet{King2003}}

\tablenotetext{f}{\citet{Roeser2011}}
\end{deluxetable}
}

\begin{deluxetable}{l|rrrrrr}
\setlength{\tabcolsep}{0.04in}
\tablewidth{0pt}
\tabletypesize{\scriptsize}
\tablecaption{UVWXYZ kinematics\label{tab:UVWXYZ}}
\tablehead{
  \colhead{Name}            &
  \colhead{U}               &
  \colhead{V}               &
  \colhead{W}               &
  \colhead{X}               &
  \colhead{Y}               &
  \colhead{Z}               \\
 
  \colhead{}                      &
  \colhead{(km s$^{-1}$)}         &
  \colhead{(km s$^{-1}$)}         &
  \colhead{(km s$^{-1}$)}         &
  \colhead{(pc)}                  &
  \colhead{(pc)}                  &
  \colhead{(pc)}                  \\

  \colhead{(1)}                   &
  \colhead{(2)}                   &
  \colhead{(3)}                   &
  \colhead{(4)}                   &
  \colhead{(5)}                   &
  \colhead{(6)}                   &
  \colhead{(7)}                   }

\startdata
\hline
NLTT 372          &       \ldots       &       \ldots       &       \ldots       & $-$21.86$\pm$2.73 & $+$61.00$\pm$7.61 & $-$56.25$\pm$7.01 \\
G 131-26AB        &       \ldots       &       \ldots       &       \ldots       &  $-$4.71$\pm$0.12 & $+$13.13$\pm$0.33 & $-$12.10$\pm$0.30 \\
SCR 0017-6645     & $-$11.30$\pm$ 1.12 & $-$16.71$\pm$ 0.87 &  $-$9.19$\pm$ 0.63 & $+$15.49$\pm$1.05 & $-$19.72$\pm$1.33 & $-$29.92$\pm$2.02 \\
GJ 2006A          &       \ldots       &       \ldots       &       \ldots       &  $+$4.04$\pm$0.23 &  $-$1.10$\pm$0.06 & $-$32.01$\pm$1.80 \\
GJ 2006B          &       \ldots       &       \ldots       &       \ldots       &  $+$4.05$\pm$0.23 &  $-$1.10$\pm$0.06 & $-$32.02$\pm$1.82 \\
SCR 0103-5515ABC  &       \ldots       &       \ldots       &       \ldots       & $+$10.93$\pm$0.70 & $-$19.50$\pm$1.26 & $-$41.64$\pm$2.68 \\
LP 467-16AB       & $-$12.34$\pm$ 1.16 & $-$16.58$\pm$ 1.19 &  $-$9.87$\pm$ 0.72 &  $-$9.55$\pm$0.37 & $+$11.37$\pm$0.44 & $-$16.01$\pm$0.62 \\
GJ 2022A          &  $-$7.71$\pm$ 1.36 & $-$26.52$\pm$ 1.47 & $-$13.93$\pm$ 1.02 &  $-$0.85$\pm$0.03 &  $-$4.24$\pm$0.14 & $-$24.95$\pm$0.81 \\
\enddata

\tablecomments{Six-dimensional phase space positions for the stars in this sample.}

\end{deluxetable}


\thispagestyle{empty}

\voffset080pt{
\begin{deluxetable}{llcccrccrcl}
\rotate
\setlength{\tabcolsep}{0.02in}
\tablewidth{0pt}
\tabletypesize{\small}
\tablecaption{Multiple Star parameters\label{tab:multiples}}
\tablehead{
  \colhead{Primary}         &
  \colhead{Secondary}       &
  \colhead{Separation}      &
  \colhead{Position Angle}  &
  \colhead{}                &
  \colhead{}                &
  \colhead{}                &
  \colhead{}                &
  \colhead{Period}          &
  \colhead{}                &
  \colhead{Resolving}       \\

  \colhead{Name}            &
  \colhead{Name}            &
  \colhead{(arcsec)}        &
  \colhead{($^{\circ}$)}      &
  \colhead{Ref.}             &
  \colhead{$\Delta mag$}    &
  \colhead{Filter}          &
  \colhead{Ref.}             &
  \colhead{}                &
  \colhead{Ref.}             &
  \colhead{Obs.}            \\

  \colhead{(1)}             &
  \colhead{(2)}             &
  \colhead{(3)}             &
  \colhead{(4)}             &
  \colhead{(5)}             &
  \colhead{(6)}             &
  \colhead{(7)}             &
  \colhead{(8)}             &
  \colhead{(9)}             &
  \colhead{(10)}            &
  \colhead{(11)}            }
\startdata
\hline
G 131-26 A      & G 131-26 B      &  0.111 & 169.9 & 2   & 0.46 & MKO $K_s$         & 2   & $\sim$4 y. & 2                    & AO \\
\hline
GJ 2006 A       & GJ 2006 B       & 17.9   &   4.9 & 1   & 0.46 & Johnson $V$       & 1   & \ldots  & \ldots               & VB \\
\hline
SCR~0103-5515 A & SCR~0103-5515 B &  0.249 &  61.0 & 3\tablenotemark{a}   & 0.20 & 2MASS $K_s$       & 3  & \ldots  & \ldots               & AO \\
SCR~0103-5515 AB& SCR~0103-5515 C &  1.77  & 336.1 & 3   & 4.43 & 2MASS $K_s$       & 3   & \ldots  & \ldots               & AO \\
\hline
LP 467-16 A     & LP 467-16 B     &  0.409 & 147.2 & 2   & 0.69 & MKO $K_s$         & 2   & \ldots  & \ldots               & AO \\ 
\hline
GJ 2022 AC      & GJ 2022 B       & 37.2   &  78.8 & 1\tablenotemark{a} & 1.94 & Johnson $V$       & 1\tablenotemark{a} & \ldots  & \ldots               & VB \\
GJ 2022 A       & GJ 2022 C       &  1.8   & 224.4 & 1   & 0.08 & Johnson $V$       & 1   & \ldots  & \ldots               & VB \\
\hline
LP 993-115      & LP 993-116 AB   & 44.8   &  60.9 & 1\tablenotemark{a} & 0.31 & Johnson $V$       & 1\tablenotemark{a} & \ldots  & \ldots               & VB \\
\enddata

\tablecomments{VB= Visual Binary.  SB = Spectroscopic Binary.  PB= Astrometric Binary with perturbation in CTIOPI data.  AO= Resolved by AO.  LI= Resolved by Lucky Imaging. SP= Resolved by Speckle Interferometry. FGS= Resolved by FGS Interferometry. References: (1) This work; (2) \citet{Beuzit2004}; (3) \citet{Delorme2013}; (4) \citet{Bergfors2010}; (5) \citet{Daemgen2007}; (6) \citet{Henry1997}; (7) \citet{Delfosse1999}; (8) \citet{Chauvin2004}; (9) \citet{Torres2006}; (10) \citet{Shkolnik2010}}

\tablenotetext{a}{Relative to unresolved components.}

\tablenotetext{b}{Estimate from PSF peaks.}

\tablenotetext{c}{Ambiguous sign on X-axis of solution.}

\end{deluxetable}
}

\begin{deluxetable}{l|ccccccccc}
\setlength{\tabcolsep}{0.04in}
\tablewidth{0pt}
\tabletypesize{\scriptsize}
\tablecaption{Deblending Ratios\label{tab:deblendingtable}}
\tablehead{
  \colhead{Desired}              &
  \multicolumn{9}{c}{Observed}   \\

  \colhead{}                      &
  \colhead{$\Delta V_J$}                  &
  \colhead{$\Delta R_{KC}$}               &
  \colhead{$\Delta I_{KC}$}               &
  \colhead{$\Delta g'_{APASS}$}           &
  \colhead{$\Delta r'_{APASS}$}           &
  \colhead{$\Delta i'_{APASS}$}           &
  \colhead{$\Delta J_{2MASS}$}            &
  \colhead{$\Delta H_{2MASS}$}            &
  \colhead{$\Delta K_{S 2MASS}$}          }

\startdata
\hline
$\Delta V_J$         & 1.00 & 1.06 & 1.36 & 1.03 & 1.07 & 1.28 & 1.67 & 1.73 & 1.80 \\
$\Delta R_{KC}$      & 0.94 & 1.00 & 1.11 & 1.19 & 1.03 & 1.12 & 1.41 & 1.46 & 1.53 \\
$\Delta I_{KC}$      & 0.74 & 0.90 & 1.00 & 0.97 & 1.03 & 1.10 & 1.26 & 1.31 & 1.37 \\
$\Delta g'_{APASS}$  & 0.97 & 0.84 & 1.03 & 1.00 & 1.04 & 1.24 & 1.38 & 1.38 & 1.43 \\
$\Delta r'_{APASS}$  & 0.93 & 0.97 & 0.97 & 0.96 & 1.00 & 1.18 & 1.32 & 1.31 & 1.36 \\
$\Delta i'_{APASS}$  & 0.78 & 0.89 & 0.91 & 0.81 & 0.85 & 1.00 & 1.11 & 1.10 & 1.14 \\
$\Delta J_{2MASS}$   & 0.60 & 0.71 & 0.79 & 0.72 & 0.76 & 0.90 & 1.00 & 1.03 & 1.07 \\
$\Delta H_{2MASS}$   & 0.58 & 0.68 & 0.76 & 0.72 & 0.76 & 0.91 & 0.97 & 1.00 & 1.04 \\
\enddata

\tablecomments{Deblending ratios, in the sense of $\Delta Y = n \times
  \Delta X$ ($\Delta V_J = 1.80 \Delta K_{s 2MASS}$).  Errors on the
  ratios themselves are $\pm$0.03; residuals of the fits are 0.5. $V_J$ 
  data were taken from \citet{Bessell1990a,Bessell1991}, \citet{Weis1993,Weis1996,Weis1999},
\citet{Koen2002}, and previous papers in this series.  $R_{KC} I_{KC}$ were taken from only 
the Bessell papers.  $g'r'i'$ were taken from APASS DR6.  $JHK_s$ were taken from the 2MASS 
Point Source Catalog \citep{Cutri2003}.  Parallax data 
were taken from the RECONS 25 pc Database, which is being prepared
for publication.}

\end{deluxetable}


\begin{figure}
\center
\includegraphics[angle=0,width=.5\textwidth]{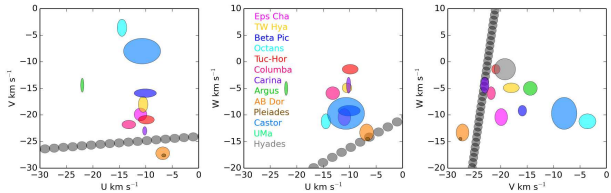}
\caption{Kinematic UVW diagrams for G 7-34.  Because we have no radial 
velocity for G 7-34, UVW velocities (gray) have been calculated for a 
range of input radial velocities, producing the ``string of pearls'' effect.  
The only possible agreement is with the AB Doradus association, with a 
best-fit radial velocity of $+$18.0 km s$^{-1}$ (which is suggestive, but 
not necessarily correct).  A similar analyses were carried out for all 
stars without known radial velocities.
\label{fig:UVW_G007-034}}
\end{figure}

\clearpage



\begin{figure} 
\centering
\includegraphics[angle=90,width=.4\textwidth]{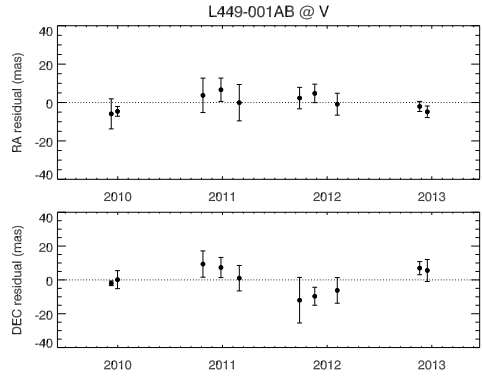}
\caption[L 449-1AB astrometric perturbation]{Nightly mean residuals of
the parallax fit (after the parallax and proper motion have been
removed) show an astrometric perturbation in both R.A. and Decl.
axes.  The CTIOPI data may cover a full orbit, but the astrometric signal is weak. 
  \label{fig:L449-001AB_orbit}}
\end{figure}

%

\end{document}